\documentclass[10pt,twocolumn,letterpaper]{article}
\usepackage{iccv}
\usepackage{times}
\usepackage{epsfig}
\usepackage{graphicx}
\usepackage{amsmath}
\usepackage{amssymb}
\usepackage{array}

\newcolumntype{C}[1]{>{\centering\let\newline\\\arraybackslash\hspace{0pt}}m{#1}}

\usepackage[pagebackref=true,breaklinks=true,letterpaper=true,colorlinks,bookmarks=false]{hyperref}
\usepackage{algorithm,algpseudocode}
\iccvfinalcopy 

\def\iccvPaperID{11457} 

\ificcvfinal\fi

\begin{document}

\title{BlindHarmony: ``Blind'' Harmonization for MR Images via Flow model}

\renewcommand{\thefootnote}{\fnsymbol{footnote}}

\author{Hwihun Jeong \qquad Heejoon Byun \qquad Dong Un Kang \qquad Jongho Lee$^\dagger$\\
Department of ECE, Seoul National University, Republic of Korea\\
{\tt\small \{hwihuni, ryanb01, qkrtnskfk23, jonghoyi\}@snu.ac.kr}
}

\maketitle
\ificcvfinal\thispagestyle{empty}\fi

\begin{abstract}
In MRI, images of the same contrast (e.g., T$_1$) from the same subject can exhibit noticeable differences when acquired using different hardware, sequences, or scan parameters. These differences in images create a domain gap that needs to be bridged by a step called image harmonization, to process the images successfully using conventional or deep learning-based image analysis (e.g., segmentation). Several methods, including deep learning-based approaches, have been proposed to achieve image harmonization. However, they often require datasets from multiple domains for deep learning training and may still be unsuccessful when applied to images from unseen domains. To address this limitation, we propose a novel concept called `Blind Harmonization', which utilizes only target domain data for training but still has the capability to harmonize images from unseen domains. For the implementation of blind harmonization, we developed BlindHarmony using an unconditional flow model trained on target domain data. The harmonized image is optimized to have a correlation with the input source domain image while ensuring that the latent vector of the flow model is close to the center of the Gaussian distribution. BlindHarmony was evaluated on both simulated and real datasets and compared to conventional methods. BlindHarmony demonstrated noticeable performance on both datasets, highlighting its potential for future use in clinical settings. The source code is available at: \url{https://github.com/SNU-LIST/BlindHarmony}
\end{abstract}

\section{Introduction}
\footnotetext{$\dagger$Corresponding author.}
Magnetic resonance imaging (MRI) is a widely-used medical imaging modality. With the advent of deep learning-based computer vision techniques, there have been numerous applications of deep learning in MRI, such as disease classification \cite{ADclass, PDclass, class3}, tumor segmentation \cite{seg1, seg2}, and solving inverse problems \cite{varnet, qsmnet}. Despite the notable performance of deep learning in MRI, its widespread use has been hindered by the inherent domain gap present in MRI data \cite{variation, variation2}. Variations occur in MRI images across different vendors, scanners, sites, and scan parameters even when the images are acquired from the same subject. This domain gap presents a generalization problem when applying the data to a neural network that has been trained on a different dataset. 

To overcome the challenges of generalization in deep learning applied to MRI data, several harmonization methods have been developed to match the source domain image to the characteristics of the target domain. These approaches include non-deep learning-based methods \cite{shinohara2017volumetric, nyul1999standardizing, shinohara2014statistical, mirzaalian2016inter, COMBAT, fortin2017harmonization} and deep learning-based methods \cite{deepharmony, cyclegan1, cyclegan2, disentangle1, taskbased1, taskbased2}, which have demonstrated performance improvements. However, there are limitations that need to be addressed. Firstly, many of these methods require multiple datasets from different domains. For instance, DeepHarmony \cite{deepharmony} which is a supervised end-to-end framework requires ``traveling subjects'' who undergo multiple MRI scans with different scanners to obtain images from both the source and target domains. Utilizing CycleGAN-based style transfer can mitigate the need for traveling subjects \cite{cyclegan1, cyclegan2}, but it still necessitates large datasets with multiple domains. Secondly, the harmonization network trained for mapping between specific source and target domains is challenging to be applied in unseen domains, limiting the generalizability of methods. Efforts have been made to employ disentanglement approaches or domain adaptation to achieve harmonization in unseen domains, but it requires a multi-contrast or multi-site paired dataset \cite{disentangle1, disentangle2}. To overcome these challenges, we propose the concept of ``Blind Harmonization'', where the harmonization network can be constructed only with the target domain data during training and applicable to diverse source domains that are unseen during training. 

\begin{figure}
\centering
  \includegraphics[width=\linewidth]{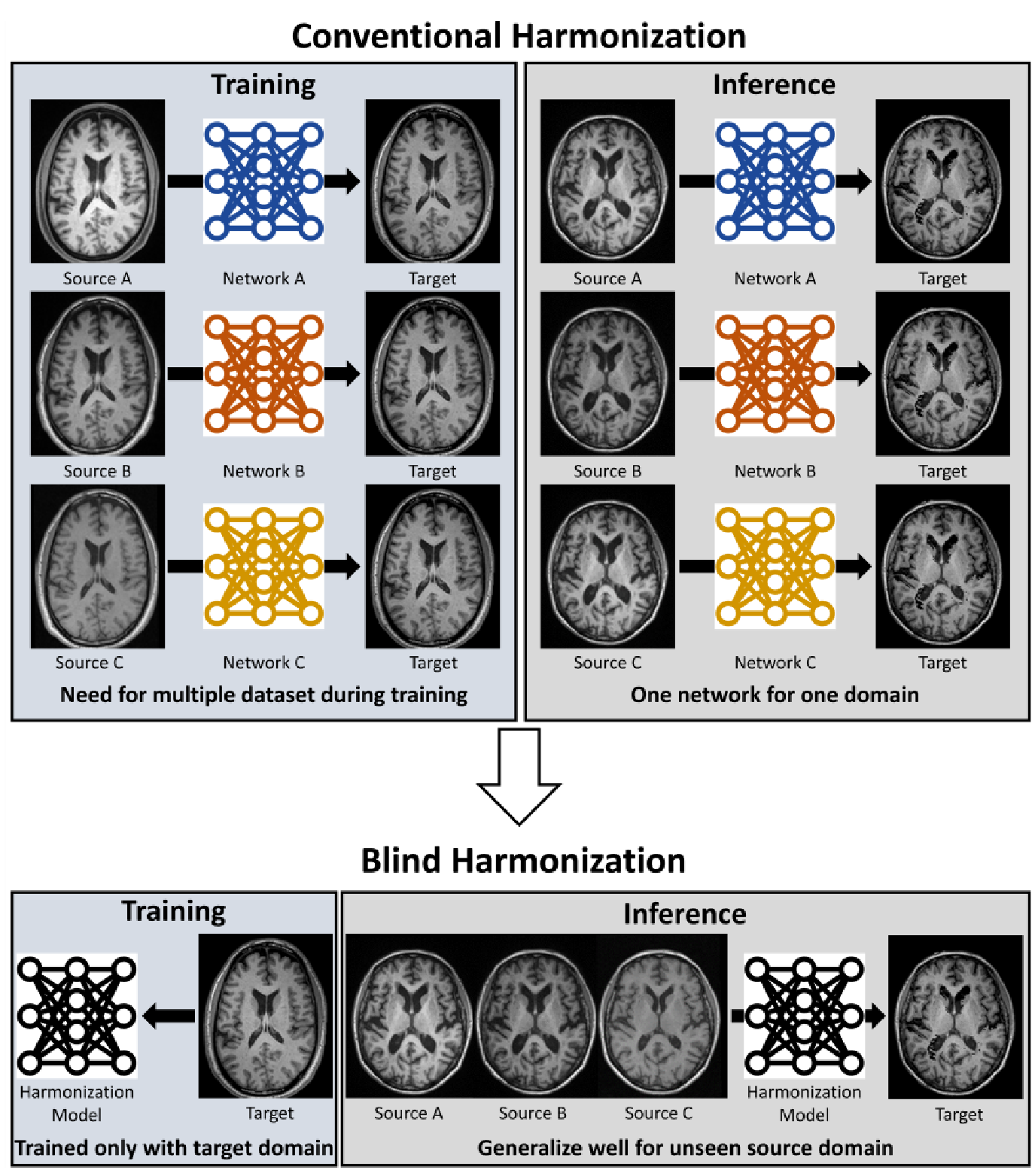}
      \caption{Blind harmonization presents an advantage over conventional harmonization models. While traditional models often necessitate multiple datasets during training or show reduced performance on unseen domains, blind harmonization can be trained solely with target domain data and generalized to previously unseen source domains.}
\label{fig:mrflowconcept}
\end{figure}

In recent years, a class of invertible generative models called normalizing flow \cite{dinh2014nice,glow,realnvp} has been introduced and has shown exceptional performance in a wide range of computer vision tasks \cite{srflow,noiseflow,flowmodel}. Normalizing flow has the unique ability not only to generate novel images that resemble samples from the distribution of a given dataset but also to map the probabilistic distribution of image datasets. This feature makes normalizing flows particularly well-suited for image generation and manipulation tasks, as they can effectively capture the underlying distribution of image data and generate new images that are consistent with that distribution. Furthermore, the invertibility of these models provides fine-grained control over the generated images, making them useful for tasks such as image manipulation and style transfer. \cite{styleflow,usman2020log}
   
In this paper, as a solution for blind harmonization, we introduce BlindHarmony which is a flow-based blind MR image harmonization framework that uses only the target domain dataset during training. Our method aims to find the harmonized image that preserves the anatomical structure and contrast of the input source domain image while maintaining a high likelihood in the flow model  (\ie, enabling harmonization for the target domain), leveraging the invertibility of flow models. Our contributions are as follows:
 \begin{enumerate}
\item We present the concept of blind harmonization, which does not require the source domain data during training and can perform harmonization for the images of untrained domains.

\item As an implementation of blind harmonization, we propose BlindHarmony. In BlindHarmony, the flow model is trained solely with the target domain data, and the harmonized image is optimized by leveraging the invertibility of the flow model.

\item We evaluate our method on both simulated data and real-world data.
\end{enumerate}

\section{Related works}
\subsection{MRI harmonization}
There are numerous demands for harmonizing MR images from different sites, vendors, or scanners. Several studies have proposed methods for harmonizing MR images from the source domain to the target domain. Conventionally, these approaches have relied on image-level post-processing techniques, such as histogram matching \cite{shinohara2014statistical,nyul2000new,nyul1999standardizing} and statistical normalization \cite{fortin2017harmonization,COMBAT,shinohara2017volumetric}, which aim to adjust the intensity values of the images to make them more similar. However, conventional methods had difficulty in capturing the subtle differences between images from different domains. For example, histogram-based models assume global histogram correspondence between images, so they ignore local contrast information. With recent advances in deep learning, there has been growing interest in developing deep learning-based methods for harmonization. One popular deep learning-based method is DeepHarmony \cite{deepharmony}, which utilizes an end-to-end supervised framework to learn the mapping between the source and target domains. Although DeepHarmony has demonstrated promising results, it requires a large dataset of traveling subjects for training, which is difficult to acquire. To address this limitation, CycleGAN-based style transfer networks \cite{cyclegan1,cyclegan2} have been employed. It can learn a mapping between images from one domain and another without the need for paired data. By training CycleGAN on a large dataset of MR images, it is possible to generate images that are visually similar to the target domain while retaining the relevant anatomical features. More recently, separated networks for the contrast network and structure network enable more flexible applicability. CALAMITI \cite{disentangle1,disentangle2} is a GAN-based method that disentangles the contrast and structural information in MR images and allows for more granular control over the image properties that need to be harmonized. In addition to image-level transformations, some works have focused on feature-level harmonization. These methods aim to learn a common feature representation that can be used for downstream analysis tasks.  For example, task-based harmonization methods \cite{taskbased1,taskbased2} learn a task-specific feature representation that can improve the performance of a specific analysis task.

\subsection{Normalizing flow} 
The normalizing flow model is a family of generative models, known as normalizing flows, that enable the parameterization of complex data distributions with a series of invertible transformations from simple random variables. In \cite{dinh2014nice}, normalizing flows were first introduced, and the NICE model was proposed as a deep learning framework that maps the complex high-dimensional density of training data to a simple factorized space using non-linear bijective transformations. Substantial improvements in invertible neural networks and high-quality image generation from the sample space have been achieved by \cite{glow,realnvp,dinh2014nice}. In particular, GLOW \cite{glow} proposed an efficient and parallelizable transformation using invertible 1$\times$1 convolutions for designing invertible neural networks and demonstrated remarkable results in high-resolution image synthesis tasks. By introducing a log-likelihood-based model in normalizing flows, GLOW can efficiently generate high-resolution natural images. Recent studies have demonstrated great performance of the normalizing flows model in a wide range of computer vision tasks such as super-resolution \cite{srflow,song2022fs}, denoising \cite{noiseflow,srflow}, and colorization \cite{li2022style} by exploiting the properties of the normalizing flows model. Among them, SRFlows \cite{srflow} adopted negative log-likelihood loss and successfully generated more diverse super-resolution images than GAN-based approaches by conditioning on low-resolution images. 

\subsection{Prior-based optimization}
Conventionally, the inverse problem of $y=Ax$ is widely solved by using regularization techniques. With the advent of generative models, several studies have proposed methods that solve this problem using generative models as prior models or regularizers. Generative adversarial networks (GANs) \cite{bora2017compressed}, normalizing flows \cite{asim2020invertible, whang2021solving}, and deep image priors \cite{ulyanov2018deep} are commonly used as priors. Individual training of these priors has the benefit of generalizability in the matrix $A$. For example, in the case of reconstructing MR images from undersampled images, a generative model prior can be applied to diverse undersampling masks \cite{jalal2021robust, luo2020mri, korkmaz2022unsupervised}.

\section{Methods}

\subsection{Harmonization model}
When a subject undergoes multiple scans with different vendors or MRI scan parameters, the resulting MR images exhibit differences, mainly in low-frequency, while the structural differences are relatively small. This provides some insight into the relationship between images from different domains. Firstly, the images are highly correlated, as the difference between domains does not largely affect the overall contrast. Secondly, the edges of the images coincide. Given $x_s$ as the source domain image and $x_h$ as its corresponding harmonized version to the target domain, the following equation holds due to the correlation and edge coincidence:
\begin{equation} \label{eq:ncc} NCC\left(x_h,x_s\right) \approx 1, \end{equation}
\begin{equation} \label{eq:grad} \| M G x_h \|_1 \approx 0. \end{equation}
In these equations, $NCC$ denotes normalized cross-correlation and $\| \|_1$ denotes the L1 norm. $M$ is a mask obtained by thresholding the gradient value of $x_s$, which retains the non-edge regions, and $G$ represents the gradient operator. Equation \ref{eq:ncc} suggests that the harmonized image should have a high cross-correlation value with the source domain image. Equation \ref{eq:grad} enforces edge sparsity in the harmonized image within regions where the source domain image is considered to have no edges (see Supplementary material for visual illustration for Eq. \ref{eq:ncc} and \ref{eq:grad}).

Based on the above formulation, we can define a distance measure, $D$, between the source domain image and the harmonized image:
\begin{equation} \label{eq:dis} D\left(x_h,x_s\right)=\beta_1\{1 - NCC\left(x_h,x_s\right)\}+\beta_2 \| M G x_h \|_1.
\end{equation}
Here $\beta$s are hyperparameters. The problem of finding $x_h$ that satisfies $D(x_h,x_s)=0$ given $x_s$ is highly ill-posed, as there exists a trivial solution of $x_h=x_s$. However, if the prior distribution of the target domain $p_X(x)$ is given, the problem can be solved using a regularization approach:
\begin{equation} \label{eq:regeq} \widehat{x_h} = \arg \min_x D\left(x,x_s\right)-\alpha \log p_X \left(x\right) ,
\end{equation}
where $\alpha$ is a regularization parameter. Equation \ref{eq:regeq} tries to generate an image that is structurally close to the source domain image while having a high probability in the target domain. The remaining issues are how to estimate the prior distribution of the target domain and how to optimize the solution for $\widehat{x_h}$. We selected the normalizing flow model to map the distribution of the target domain, because the inherent invertibility of the flow model can provide an advantage for optimization.

\subsection{Flow-based prior learning}
A normalizing flow is an invertible transformation that maps a sample from a simple probability distribution (\eg, normal Gaussian) to a sample from a complex probability distribution. The transformation itself (often called ``flow'') and its inverse are assumed to be differentiable.

Let $Z \in \mathbb{R}^D$ be a random variable with an associated probability density function (PDF) $p_Z: \mathbb{R}^D \rightarrow [0,1]$ which is assumed to be known and tractable. Let $f_\theta : \mathbb{R}^D \rightarrow \mathbb{R}^D$ be a diffeomorphism parameterized by the vector $\theta \in \mathbb{R}^P$ with an inverse denoted by $f_\theta^{-1}$. Then the PDF of the random variable $X=f_\theta^{-1}\left(Z\right)$ can be computed explicitly using the change of variables formula: \begin{equation} \label{eq:flow} p_X\left(x|\theta\right) = p_Z\left(f_\theta\left(x\right)\right) 
   \left| \det\frac{\partial
   {f_\theta\left(x\right)}}{\partial{x}} \right|, \end{equation}
where $\frac{\partial{f_\theta\left(x\right)}}{\partial{x}}$ is the Jacobian of $f_\theta$.

When applying normalizing flows for sample generation or density estimation problems, the simple distribution $p_Z$, known as the ``latent distribution'', is transformed via the ``flow''  $f_\theta$ to a more complex distribution $p_X$. The objective for both problems is to find the value of the parameters $\theta$ for which $p_X$ closely approximates the underlying distribution $p_{data}$ of the given dataset. Only after the objective is satisfied can we accurately estimate the densities of the random samples using the change of variable formula or generate random samples that are consistent with the given data by first sampling from the latent distribution and feeding the sample to the inverse of the flow.

The aforementioned objective can be stated formally as a maximum likelihood estimation (MLE) problem: maximizing the expected log-likelihood
\begin{align} & \mathcal{L}\left(\theta;x\right) :=E_{X \sim p_{data}}[\log\left(p_X\left(X|\theta\right)\right)] \\
\label{eq:mle}   &\approx \frac{1}{N} \sum\limits_{i=1}^{N} \log \left(p_X\left(x^{\left(i\right)}|\theta\right)\right) \\
  &= \frac{1}{N} \sum\limits_{i=1}^{N} \left[\log\left(p_Z\left(f_\theta\left(x^{\left(i\right)}\right)\right)\right)+ 
   \log \left| \det\frac{\partial{f_\theta\left(x^{(i)}\right)}}{\partial{x}} \right|  \right]. \end{align} 
over the possible values of the parameters $\theta \in \mathbb{R}^P$, where $\mathcal{D} := \{x^{\left(i\right)}\}_{i=1}^N $ is the given dataset. Therefore, training the normalizing flow involves updating the parameters of the flow so that the expected log-likelihood is maximized.

In order to accurately and efficiently approximate the target distribution $p_{data}$, a normalizing flow $f_\theta$ must satisfy several conditions: It must be a bijection with differentiable forward and inverse transformations, it must be expressive enough to model the complexity of the target distribution, and the computations of $f_\theta$, $f_\theta^{-1}$, and $\det\frac{\partial{f_\theta\left(x\right)}}{\partial{x}}$ must be done efficiently.

Therefore, many state-of-the-art normalizing flows use neural networks that are carefully designed to have differentiable inverse transformations and a Jacobian matrix whose determinant can be computed efficiently. These include coupling transforms, which have been shown to be particularly effective.

\begin{figure}
\centering
      \includegraphics[width=\linewidth]{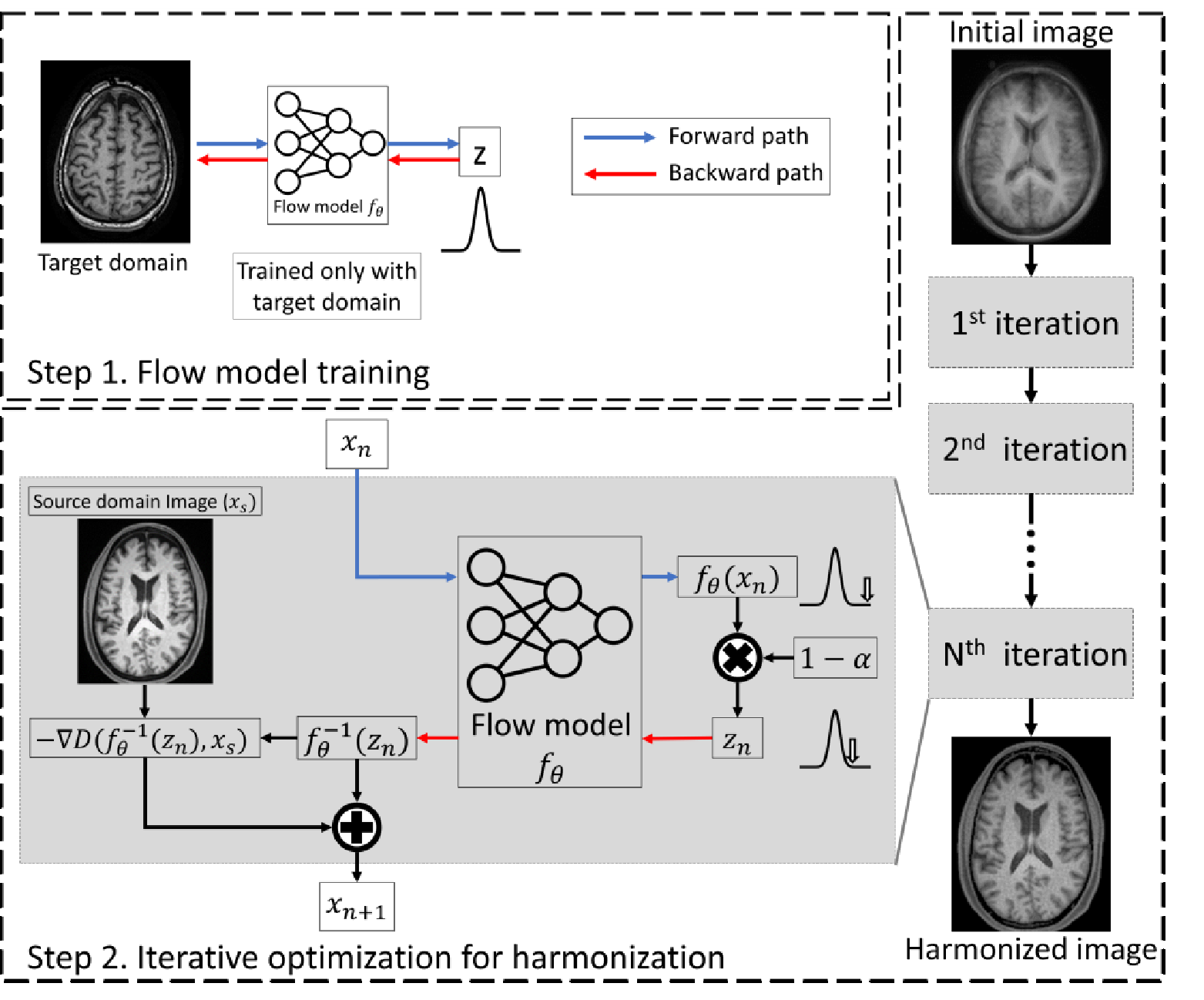}
      \caption{The BlindHarmony framework operates as follows: first, a flow model is trained solely on target domain data. Then, harmonization is performed iteratively on both latent variable and image domains using the flow model.}
   \label{fig:mrflowexplain}
\end{figure}
   
\subsection{BlindHarmony optimization}
In order to harmonize images from unknown domains, an unconditional flow model is trained on the target domain only. The prior distribution of the harmonized image $x$ can be parameterized as follows:
 \begin{equation} \label{eq:harmodel2} \log p_X\left(x\right)= \log \left(p_Z\left(z\right)\right) + \log \left|\det\frac{\partial
   {f_\theta\left(x\right)}}{\partial{x}}\right|. \end{equation}
If $z$ is a normal Gaussian, we can rewrite Equation \ref{eq:regeq} in the $z$-domain as follows:
\begin{equation} \label{eq:harmodel3} \widehat{z_h} = \arg \min_z D\left(f_\theta^{-1}\left(z\right),x_s\right)-\alpha \log p_Z \left(z\right) \end{equation}
\begin{equation} \label{eq:harmodel4}= \arg \min_z D\left(f_\theta^{-1}\left(z\right),x_s\right) + \alpha |z|^2. \end{equation} 

The optimization process of Equation \ref{eq:harmodel4} requires the calculation of gradients of $f_\theta^{-1}\left(z\right)$, which can be computationally burdensome. To increase computational efficiency and reduce processing time, we simply omit the calculation of $\frac{\partial{f_\theta^{-1}\left(z\right)}}{\partial{z}}$. Instead, iterative optimization is performed in both the $z$- and $x$-domains, by leveraging the invertibility of normalizing flow. The algorithm alternates between a gradient descent of the distance measure $D\left(x,x_s\right)$ and a gradient descent of the prior term $|z|^2$. 

In the latent vector domain $z$, $z$ is updated so that it does not deviate far from the center of the Gaussian:
\begin{equation} \label{alpha} z \rightarrow (1-\alpha)z. \end{equation}
In the image domain $x$, the gradient of $D\left(x,x_s\right)$ is measured and updated at each iteration as follows:
\begin{equation} \label{beta} x \rightarrow x + \nabla_x[\beta_1 NCC\left(x,x_s\right) 
   - \beta_2 \| M G x_h \|_1]. \end{equation}
After $N$ iterations, the resultant image $x$ is a harmonized image $\widehat{x_h}$. The overall algorithm is formulated as Algorithm \ref{al:BlindHarmony} (Fig. \ref{fig:mrflowexplain}). The hyperparameters $\alpha$, $\beta_1$, and $\beta_2$ are found heuristically using grid search. The hyperparameters have been fixed as: $\alpha = 0.001,\beta_1 = 1000,\beta_2 = 0.001,$ and $N = 10$. The initial image $x_0$ is chosen to be the averaged image of the training data images.

\begin{algorithm}
   \caption{BlindHarmony optimization}
   $x_s$: Source domain image
   
   $\widehat{x_h}$: Harmonized image

   $f_\theta$: Flow model trained on the target domain
   
   $x_0$: Initial image
   
   $\alpha$, $\beta_1$, $\beta_2$: Hyperparameters

   $N$: The number of iteration
   
   \begin{algorithmic}[1]
   \Require{$x_s$, $x_0$, $\alpha$, $\beta_1$, $\beta_2$}
   \For{$n = 0,1,\ldots,N-1$}
   \State{$x_{n+1} = f_\theta^{-1}\left(z_n\right) + \nabla_x[\beta_1 NCC\left(x,x_s\right) 
   - \beta_2 \| M G x_h \|_1]_{x=f_\theta^{-1}\left(z_n\right)}$}
   \State{$z_{n+1} = \left(1-\alpha\right)f_\theta(x_{n+1})$}
   \EndFor
   \State{$\widehat{x_h} = x_N$}
   \end{algorithmic}
   \label{al:BlindHarmony}
\end{algorithm}

\section{Experiments} 
\subsection{Dataset}
T$_1$-weighted images in the OASIS3 dataset \cite{lamontagne2019oasis} were used to train and evaluate the proposed framework. The OASIS3 dataset consists of images scanned with different scanners. Images acquired with the Siemens TrioTim scanner were used as the target domain. For the source domain datasets, three datasets consisting of images acquired with different manufacturer models (Domain 1,2,3) and a dataset from a different scanner with the same manufacturer model (Domain 4) were used (see Supplementary material). All images were resampled to the same resolution of 1.2$\times$1.2$\times$1.2 mm$^3$ and min-max normalized in a slice level. 

\subsection{Network training detail}
In our experiments, we used the Neural Spline Flow (NSF) architecture with rational quadratic (RQ) spline coupling layers that was outlined by \cite{durkan2019neural}. A Glow-like multiscale architecture was used, following NSF (Durkan et al., 2019) and Glow \cite{glow}. Each layer of the network contains 7 transformation steps, where each step consists of an actnorm layer, an invertible 1$\times$1 convolution an RQ spline coupling transform, and another 1$\times$1  convolution. The network consists of 4 layers, which results in a total of 28 coupling transformation steps. Also, 3 residual blocks and batch normalization layers are included in the subnetworks parameterizing the RQ splines. An Adam optimizer with an initial learning rate of 0.0005 and cosine annealing of the learning rate was used to iteratively optimize the parameters up to 20K steps. The sampled images are reported in the Supplementary material.

\subsection{Simulated data evaluation}
To evaluate the effectiveness of our proposed harmonization approach, we developed simulated data by applying three different image transformations: exponential transformation (Domain-Exp), log transformation (Domain-Log), and Gamma transformations with powers of 0.7 (Domain-Gamma0.7) to the target domain images. The target domain images were normalized with min-max normalization, then the three above transformations were applied. The min-max normalization was performed again on the transformed images to generate simulated source domain data. We applied BlindHarmony to the source domain.

To evaluate the performance of our proposed method, we compared it with the following methods: slice-wise histogram matching (HM), low-frequency replacing (SSIMH) \cite{guan2022fast}, end-to-end U-net \cite{ronneberger2015u}, and unsupervised CycleGAN \cite{zhu2017unpaired}. We trained U-net$_{Exp}$, U-net$_{Log}$, U-net$_{Gamma0.7}$, CycleGAN$_{Exp}$, CycleGAN$_{Log}$, and CycleGAN$_{Gamma0.7}$ models for each domain mapping and used them for comparison (\eg, U-net$_{Exp}$ was trained on Domain-Exp data).

\begin{figure*}
\setlength{\tabcolsep}{0pt}
\renewcommand{\arraystretch}{0.5}
\centering
   {\footnotesize
\begin{tabular}{C{1.5cm}C{2.3cm}C{2.3cm}C{2.3cm}C{2.3cm}C{2.3cm}C{2.3cm}C{2.3cm}}
&Source & Target & BlindHarmony & HM & SSIMH\cite{guan2022fast} & CycleGAN$_{Exp}$\cite{zhu2017unpaired} &U-net$_{Exp}$\cite{ronneberger2015u}\\
        Domain-Exp&
         \includegraphics[width=0.13\textwidth]{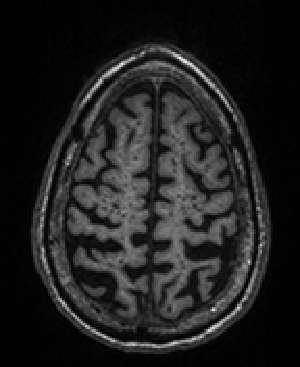}&
         \includegraphics[width=0.13\textwidth]{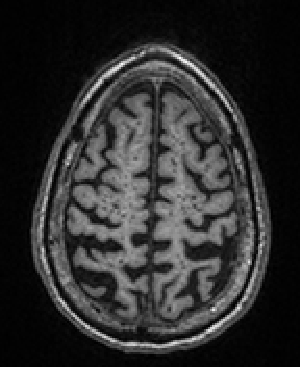}&
         \includegraphics[width=0.13\textwidth]{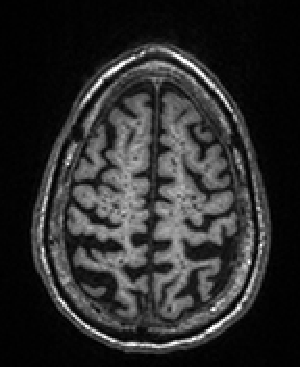}&
         \includegraphics[width=0.13\textwidth]{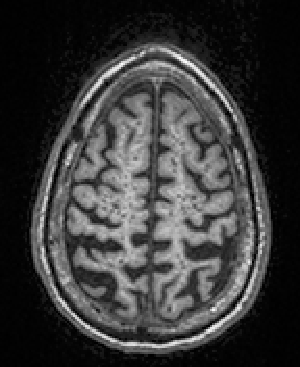}&
         \includegraphics[width=0.13\textwidth]{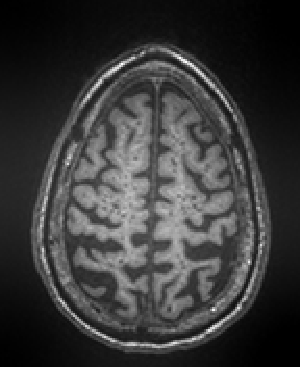}&
         \includegraphics[width=0.13\textwidth]{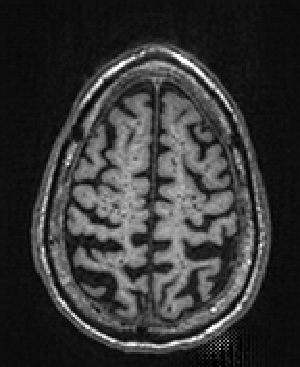}&
         \includegraphics[width=0.13\textwidth]{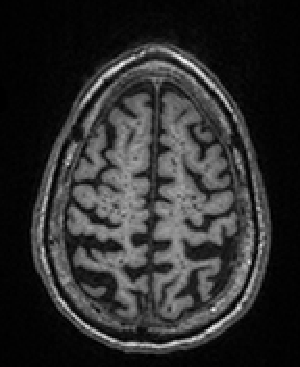}\\
         
         Domain-Log&
         \includegraphics[width=0.13\textwidth]{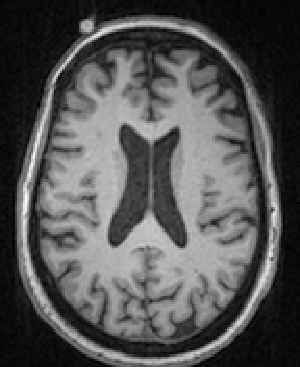}&
         \includegraphics[width=0.13\textwidth]{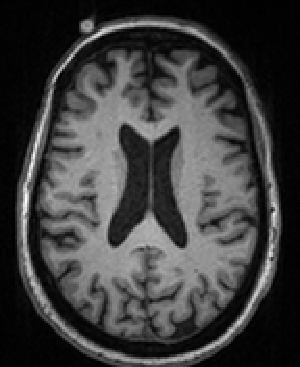}&
         \includegraphics[width=0.13\textwidth]{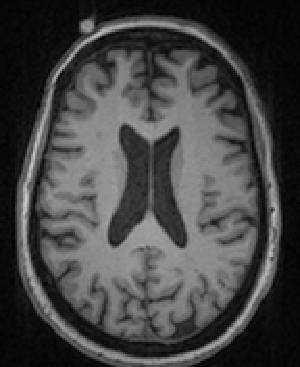}&
         \includegraphics[width=0.13\textwidth]{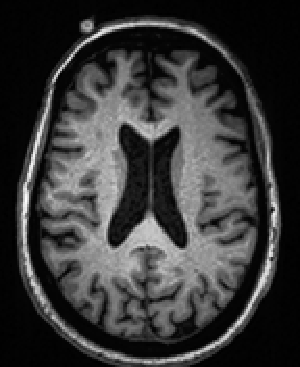}&
         \includegraphics[width=0.13\textwidth]{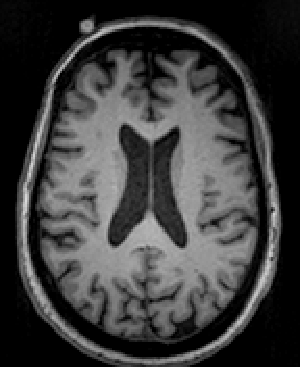}&
         \includegraphics[width=0.13\textwidth]{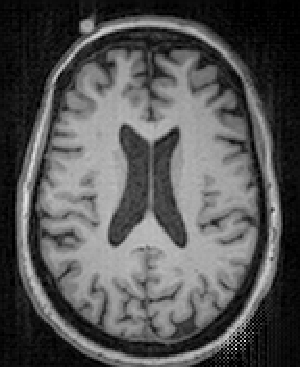}&
         \includegraphics[width=0.13\textwidth]{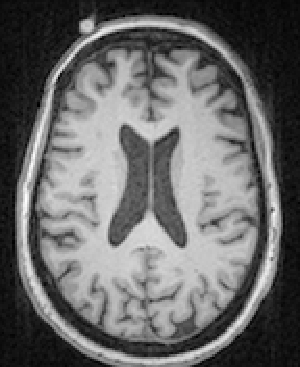}\\
         
         Domain-Gamma0.7&
         \includegraphics[width=0.13\textwidth]{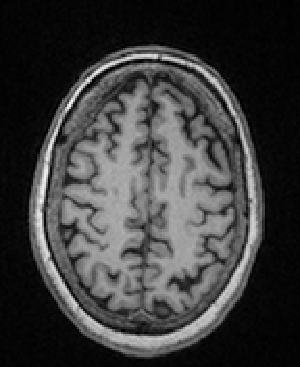}&
         \includegraphics[width=0.13\textwidth]{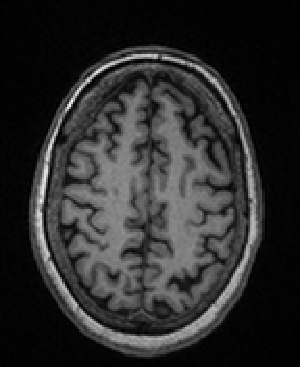}&
         \includegraphics[width=0.13\textwidth]{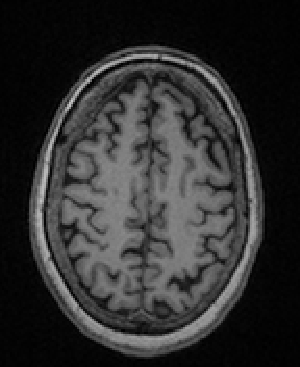}&
         \includegraphics[width=0.13\textwidth]{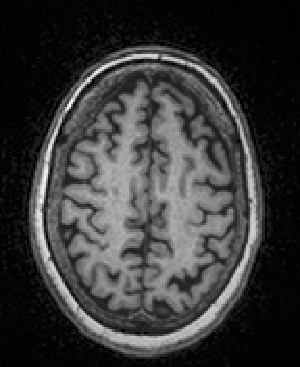}&
         \includegraphics[width=0.13\textwidth]{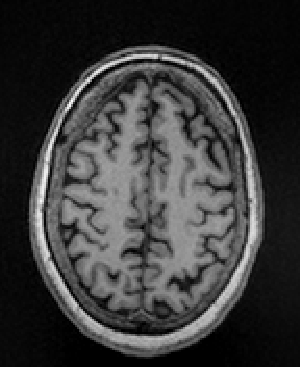}&
         \includegraphics[width=0.13\textwidth]{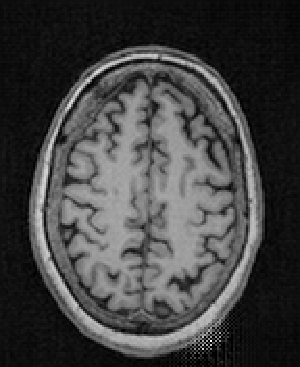}&
         \includegraphics[width=0.13\textwidth]{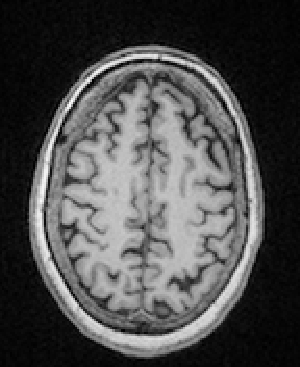}\\
     \end{tabular}
     }
     \caption{Example image of BlindHarmony application of simulated source domain images and comparison with other harmonization methods.}
   \label{fig:retro}
 \end{figure*}

Figure \ref{fig:retro} displays the results of our harmonization approach using BlindHarmony on each of the simulated source domain images (1st column), along with the target domain image (2nd column), and the other methods. A visual inspection of the results confirms the effectiveness of our approach in harmonizing images. On the other hand, CycleGAN and U-net fail to harmonize a source domain image when trained on another source domain dataset (\eg, U-net$_{Exp}$ which was trained on Domain-Exp data while applying Domain-Gamma0.7 data). In contrast, BlindHarmony offers a more efficient and versatile solution by utilizing a single network for harmonization across diverse source domains.

\begin{table*}[]
\centering
   {\footnotesize
   \begin{tabular}{ccccccccc}
   \hline
               & \multicolumn{2}{c}{}& \multicolumn{2}{c}{Domain-Exp} & \multicolumn{2}{c}{Domain-Log} & \multicolumn{2}{c}{Domain-Gamma0.7}\\ 
              & Unsupervised & Blind  & PSNR($\uparrow$) & SSIM($\uparrow$) & PSNR($\uparrow$) & SSIM($\uparrow$) & PSNR($\uparrow$) & SSIM  ($\uparrow$)\\
              \hline \hline
   Source                &  &   & 22.6     & 0.952   & 21.4    & 0.958      & 21.6       & 0.955          \\
   \hline
   BlindHarmony (Ours)   &O & O & \textbf{29.6} & \textbf{0.985} & \textbf{28.8} & \textbf{0.978} & \textbf{27.4} & 0.969 \\
   HM                    &O & O & 26.5          & 0.961          & 26.5          & 0.961          & 26.5          & 0.961          \\
   SSIMH\cite{guan2022fast}                  &O & O & 26.5          & 0.972          & 25.8          & 0.973          & 26.3          & \textbf{0.976}          \\
   \hline
   CycleGAN$_{Exp}$\cite{zhu2017unpaired}      &O & X & \textbf{32.6} & \textbf{0.993} & 23.0          & 0.951          & 23.0          & 0.948          \\
   CycleGAN$_{Log}$      &O & X & 22.8          & 0.943          & 35.5          & \textbf{0.996} & 34.5          & 0.995          \\
   CycleGAN$_{Gamma0.7}$  &O & X & 22.1          & 0.932          & \textbf{35.6} & 0.996         & \textbf{35.6} & \textbf{0.996} \\
   \hline
   U-net$_{Exp}$\cite{ronneberger2015u}         &X & X & \textbf{65.6} & \textbf{0.999} & 15.9          & 0.885          & 15.9          & 0.879          \\ 
   U-net$_{Log}$         &X & X & 16.8          & 0.803          &\textbf{56.5}  & \textbf{0.999} & 38.0          & 0.997          \\ 
   U-net$_{Gamma0.7}$     &X & X & 16.3          & 0.766          & 39.2          & 0.998          & \textbf{55.1} & \textbf{0.999} \\ \hline
   \end{tabular}
   }
   
   \caption{The PSNR and SSIM values calculated between the harmonized image from the simulated source domain and the reference of the target domain image. The regions with signals were used as a mask.}
   \label{table:retro}
\end{table*}

Table \ref{table:retro} presents the results of our simulated data evaluation of the harmonization methods. We calculated the peak signal-to-noise ratio (PSNR) and structural similarity (SSIM) values using the target image as a reference. The table shows that our proposed BlindHarmony framework outperforms the source domain image, as evidenced by the improved PSNR and SSIM values. The averaged PSNR value improved from 21.9 dB for the source domain images to 28.6 dB for the BlindHarmony harmonized images. These results demonstrate the effectiveness of BlindHarmony in harmonizing images from different domains.

It is worth noting that U-net and CycleGAN outperformed BlindHarmony when they were trained separately for each source domain (\eg, U-net$_{Exp}$ which was trained on Domain-Exp data while applying Domain-Exp data). However, BlindHarmony used only one network for all source domains, making it a practical solution for harmonizing images from multiple source domains.

\begin{figure*}
\setlength{\tabcolsep}{0pt}
\renewcommand{\arraystretch}{0.5}
\centering
   {\footnotesize
\begin{tabular}{C{1.3cm}C{2.3cm}C{2.3cm}C{2.3cm}C{2.3cm}C{2.3cm}C{2.3cm}C{2.3cm}}
&Source & Target & BlindHarmony & HM & SSIMH\cite{guan2022fast} & CycleGAN$_{each}$\cite{zhu2017unpaired} &U-net$_{each}$\cite{ronneberger2015u}\\
        Domain 1&
         \includegraphics[width=0.13\textwidth]{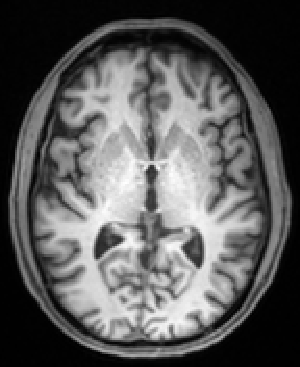}&
         \includegraphics[width=0.13\textwidth]{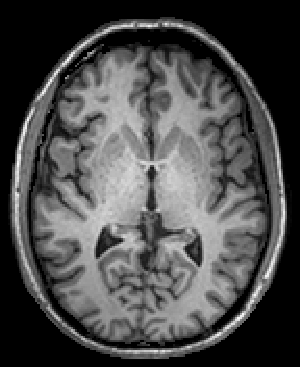}&
         \includegraphics[width=0.13\textwidth]{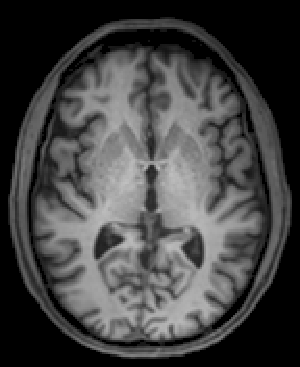}&
         \includegraphics[width=0.13\textwidth]{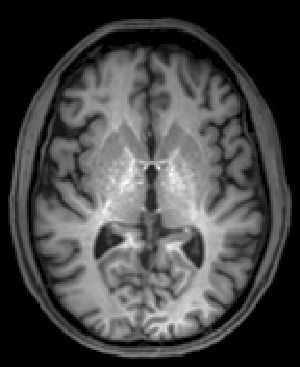}&
         \includegraphics[width=0.13\textwidth]{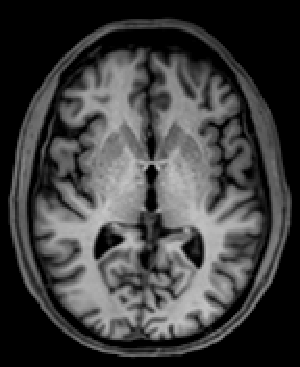}&
         \includegraphics[width=0.13\textwidth]{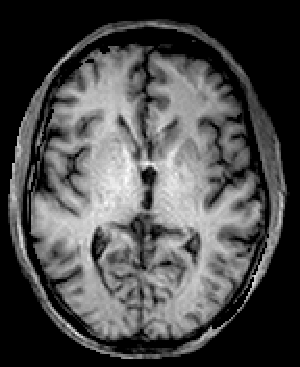}&
         \includegraphics[width=0.13\textwidth]{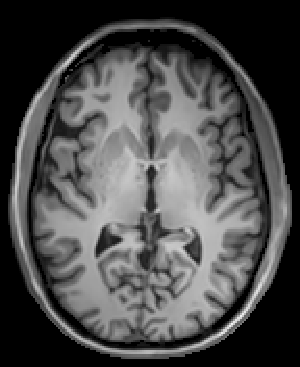}\\
         
         Domain 2&
         \includegraphics[width=0.13\textwidth]{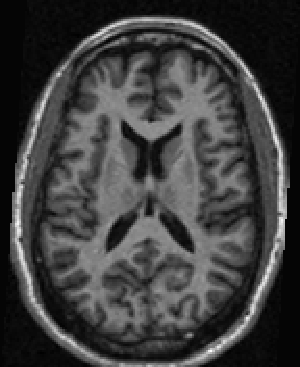}&
         \includegraphics[width=0.13\textwidth]{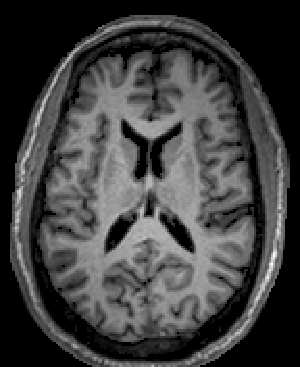}&
         \includegraphics[width=0.13\textwidth]{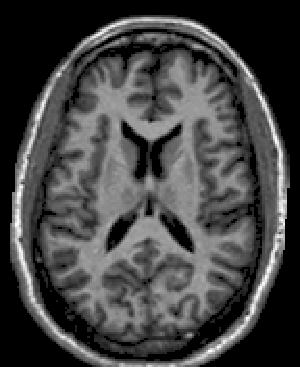}&
         \includegraphics[width=0.13\textwidth]{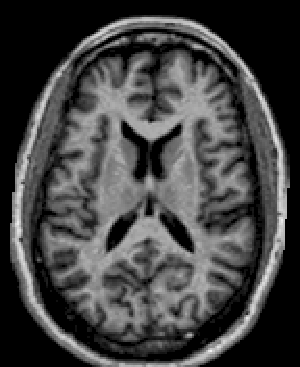}&
         \includegraphics[width=0.13\textwidth]{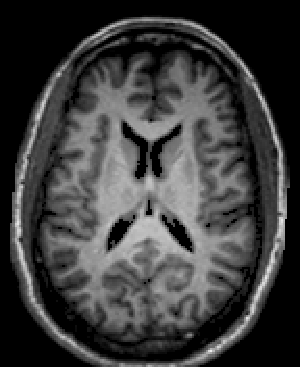}&
         \includegraphics[width=0.13\textwidth]{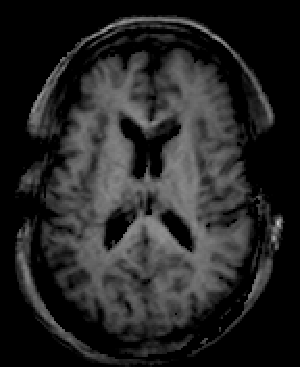}&
         \includegraphics[width=0.13\textwidth]{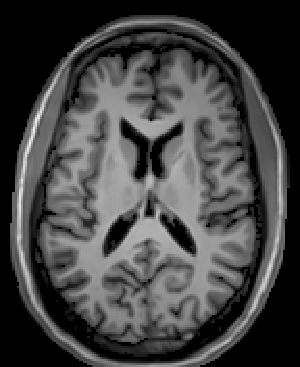}\\
         
         Domain 3&
         \includegraphics[width=0.13\textwidth]{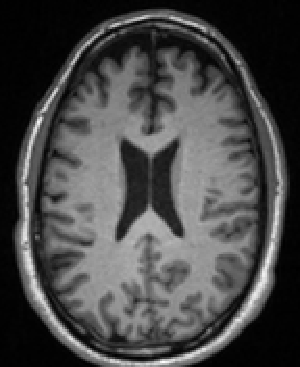}&
         \includegraphics[width=0.13\textwidth]{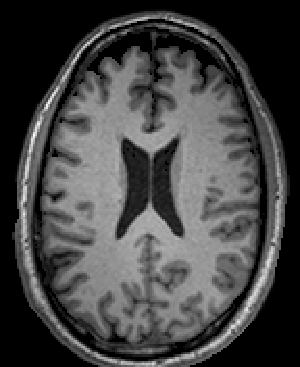}&
         \includegraphics[width=0.13\textwidth]{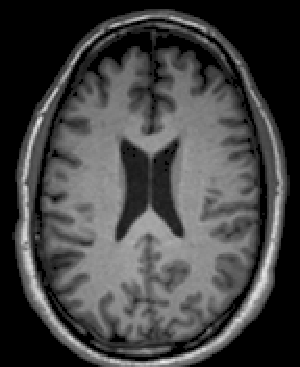}&
         \includegraphics[width=0.13\textwidth]{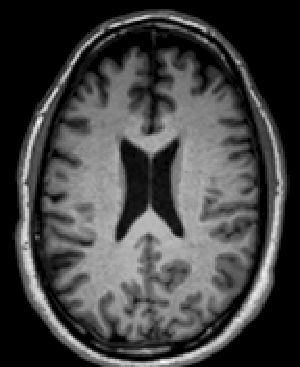}&
         \includegraphics[width=0.13\textwidth]{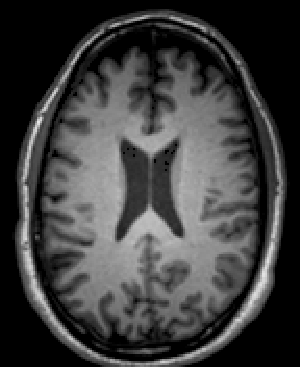}&
         \includegraphics[width=0.13\textwidth]{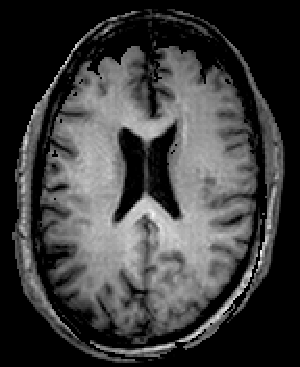}&
         \includegraphics[width=0.13\textwidth]{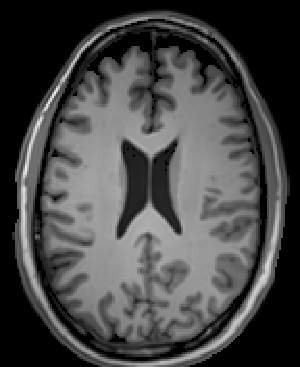}\\
         
         Domain 4&
         \includegraphics[width=0.13\textwidth]{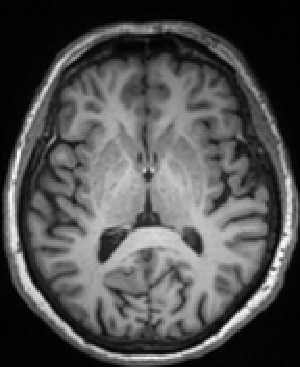}&
         \includegraphics[width=0.13\textwidth]{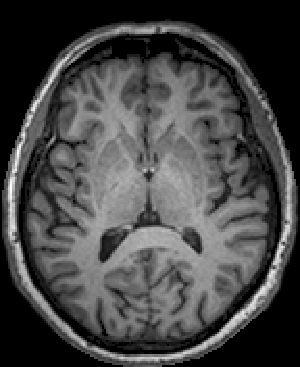}&
         \includegraphics[width=0.13\textwidth]{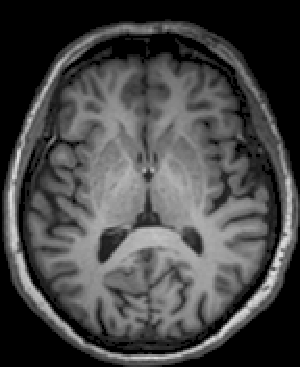}&
         \includegraphics[width=0.13\textwidth]{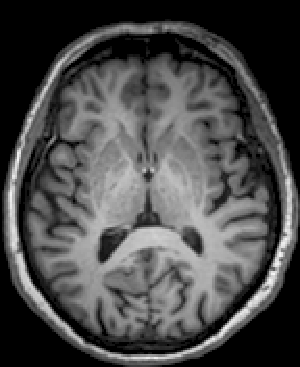}&
         \includegraphics[width=0.13\textwidth]{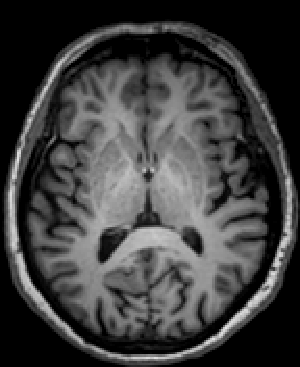}&
         \includegraphics[width=0.13\textwidth]{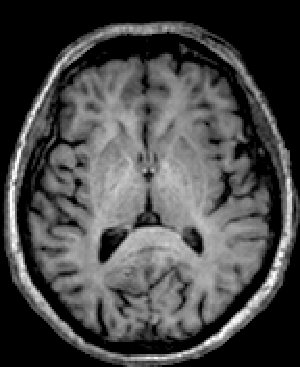}&
         \includegraphics[width=0.13\textwidth]{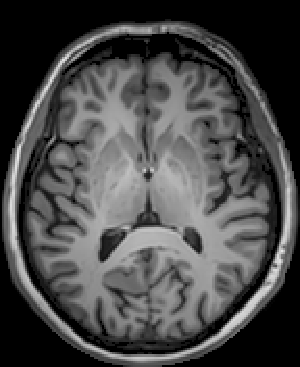}\\

     \end{tabular}
     }
      \caption{Example image of BlindHarmony application of real source domain images and comparison
      with other harmonization methods. CycleGAN and U-net results stand for the output of the network that was trained for each source domain.}
   \label{fig:pro}
 \end{figure*}

\subsection{Real-world data application}
In addition to the simulation dataset, we also evaluated BlindHarmony on four real datasets from different scanners. Twenty traveling subjects from OASIS 3 \cite{lamontagne2019oasis} dataset who underwent multi-scanner scans were utilized in order to compare the results quantitatively. The image of each source domain is registered to the target domain image by using the FSL FLIRT \cite{jenkinson2002improved} function. We compared the results with the other methods (U-Net and CycleGAN retrained for these datasets and conventional methods of volumetric HM and SSIMH). Figure \ref{fig:pro} presents the results of harmonizing the source domain images (column 1) to the target domain images (column 2) using BlindHarmony (column 3). BlindHarmony effectively harmonized the images and reduced the inter-scanner variability, bringing them closer to the target domain images. The BlindHarmony also demonstrated superior harmonization performance not only to the conventional methods (SSIMH and HM) but also to the CycleGAN, which illustrates structural distortion.

\begin{table*}[]
\centering
      {\footnotesize
   \begin{tabular}{ccccccccccc}
   \hline
            & \multicolumn{2}{c}{}& \multicolumn{2}{c}{Domain1} & \multicolumn{2}{c}{Domain2} & \multicolumn{2}{c}{Domain3} & \multicolumn{2}{c}{Domain4}\\ 
            & Unsupervised & Blind & PSNR($\uparrow$) & SSIM($\uparrow$) & PSNR($\uparrow$) & SSIM($\uparrow$) & PSNR($\uparrow$) & SSIM($\uparrow$) & PSNR($\uparrow$) & SSIM  ($\uparrow$) \\
            \hline \hline
   Source            &  &   & 19.6          & 0.833 & 19.4    & 0.836      & 23.0       & 0.893      & 24.1    & 0.914    \\
   \hline
   BlindHarmony (Ours)     &O & O & 20.2          & \textbf{0.840} & \textbf{20.8}       & \textbf{0.850}      & 23.0 & 0.892      & \textbf{24.6}    & 0.912     \\
   HM                &O & O & 20.4          & 0.834 & 20.6       & 0.840      & 22.5       & 0.882      & 23.9    & 0.899   \\
   SSIMH\cite{guan2022fast}              &O & O & \textbf{20.4}          & 0.831 & 20.4       & 0.833      & 22.0       & 0.882      & 22.6    & 0.896     \\ 
   \hline
   CycleGAN$_{each}$\cite{zhu2017unpaired} &O & X & 7.22          & 0.451 & 15.3       & 0.612      & 6.62       & 0.442      & 19.8    & 0.795    \\
   U-net$_{each}$\cite{ronneberger2015u}    &X & X & 25.0 & 0.919 & 23.4 & 0.890      & 25.1       & 0.925      & 25.6    & 0.920  \\ \hline
   \end{tabular}}
   \caption{The quantitative results of application to real-world data. PSNR and SSIM values were calculated by using the target domain image as a reference. The case using BlindHarmony illustrated improved consistency to the target domain image. The regions with signals were used as a mask. CycleGAN$_{each}$ and U-net$_{each}$ stand for CycleGAN and U-net trained on each source domain (\eg, CycleGAN$_{each}$ for Domain 1 application is trained on Domain 1 dataset).}
   \label{table:pro}
\end{table*}

The quantitative evaluation using PSNR and SSIM metrics demonstrated improvements in both PSNR and SSIM values compared to the source images (averaged PSNR: 21.5 dB to 22.2 dB). In particular, BlindHarmony has exhibited superior metric results compared to the HM, SSIMH, and CycleGAN algorithms.

The effect size of PSNR and SSIM improvement observed in this study is smaller than that in the simulated data study. This can be attributed to two key factors. Firstly, the domain gap between the source and target domains might be smaller than that in the simulated data, leading to a weaker harmonization effect. Secondly, the registration process between the source and target domains may not be perfectly aligned due to potential errors in registration and the time gap between separate scans. These factors may have led to a reduced effect of harmonization of effect of harmonization in the metric calculation.

\begin{figure*}
\setlength{\tabcolsep}{0pt}
\renewcommand{\arraystretch}{0.5}
\centering
   {\scriptsize
\begin{tabular}{C{1cm}C{2.05cm}C{2.05cm}C{2.05cm}C{2.05cm}C{2.05cm}C{2.05cm}C{2.05cm}C{2.05cm}}
&Source & Label & No harmonization & BlindHarmony & HM & SSIMH\cite{guan2022fast} & CycleGAN$_{each}$\cite{zhu2017unpaired} &U-net$_{each}$\cite{ronneberger2015u}\\
        Domain 1&
         \includegraphics[width=0.116\textwidth]{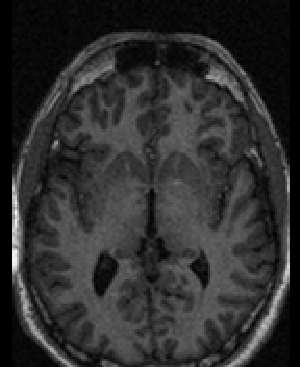}&
         \includegraphics[width=0.116\textwidth]{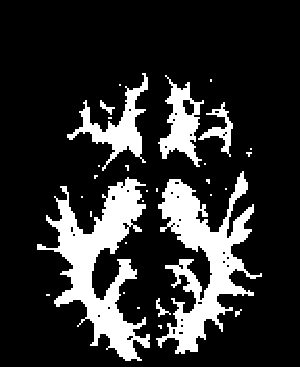}&
         \includegraphics[width=0.116\textwidth]{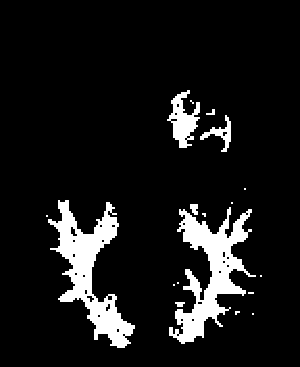}&
         \includegraphics[width=0.116\textwidth]{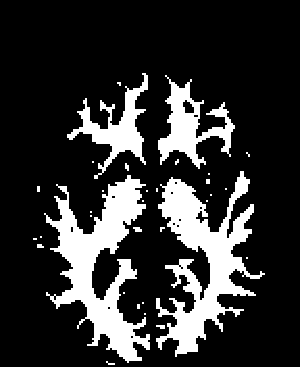}&
         \includegraphics[width=0.116\textwidth]{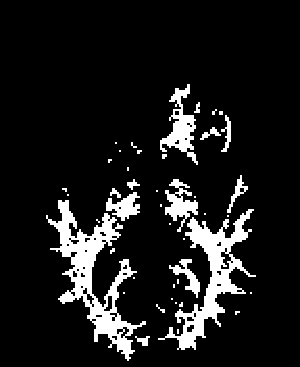}&
         \includegraphics[width=0.116\textwidth]{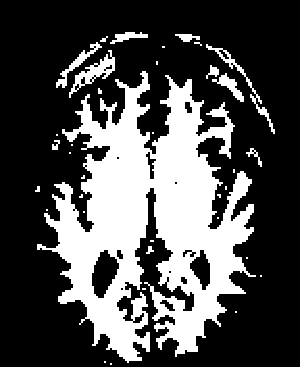}&
         \includegraphics[width=0.116\textwidth]{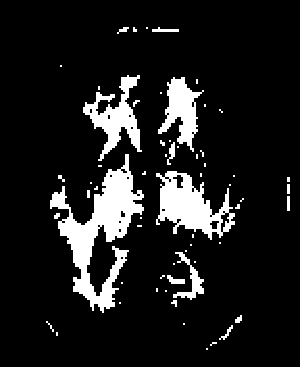}&
         \includegraphics[width=0.116\textwidth]{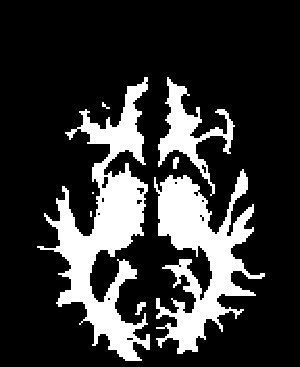}\\
         
         Domain 2&
         \includegraphics[width=0.116\textwidth]{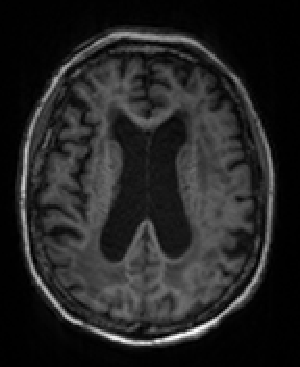}&
         \includegraphics[width=0.116\textwidth]{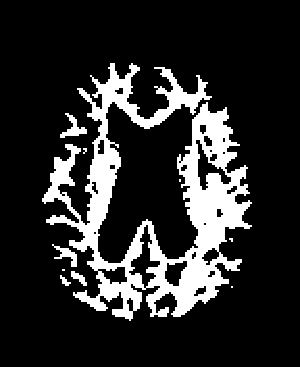}&
         \includegraphics[width=0.116\textwidth]{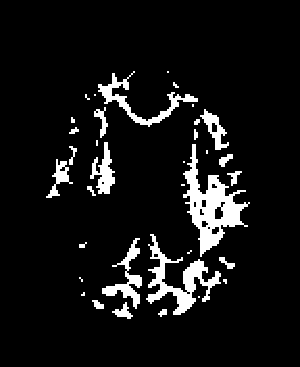}&
         \includegraphics[width=0.116\textwidth]{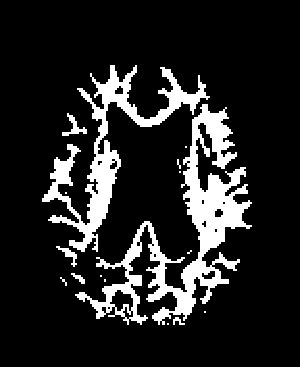}&
         \includegraphics[width=0.116\textwidth]{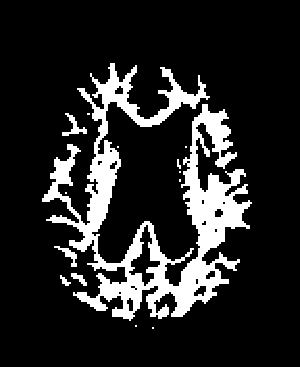}&
         \includegraphics[width=0.116\textwidth]{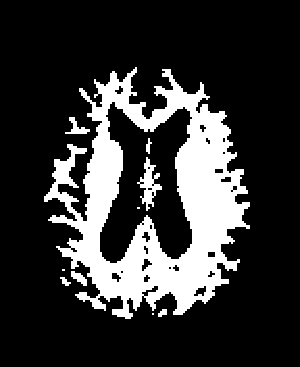}&
         \includegraphics[width=0.116\textwidth]{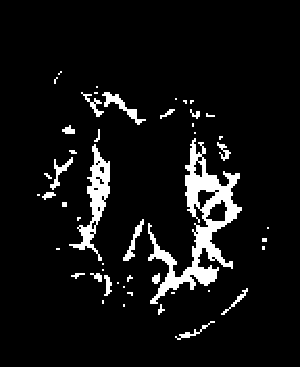}&
         \includegraphics[width=0.116\textwidth]{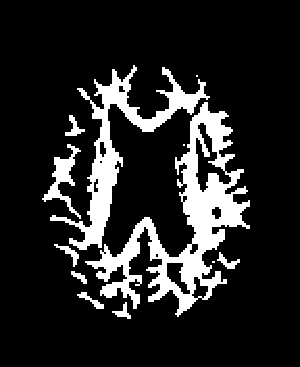}\\
         
        Domain 3&
         \includegraphics[width=0.116\textwidth]{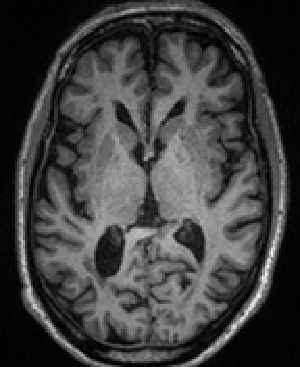}&
         \includegraphics[width=0.116\textwidth]{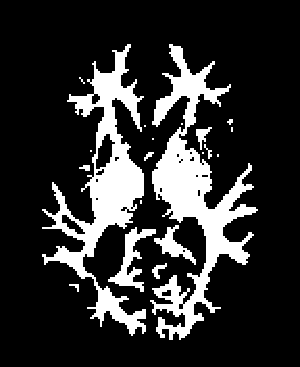}&
         \includegraphics[width=0.116\textwidth]{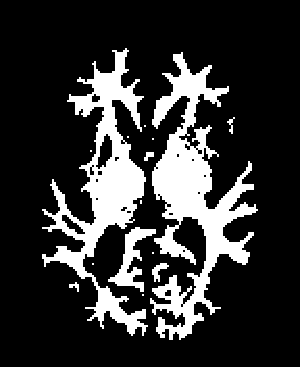}&
         \includegraphics[width=0.116\textwidth]{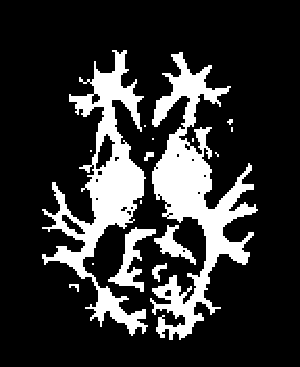}&
         \includegraphics[width=0.116\textwidth]{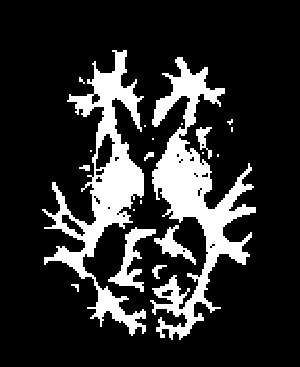}&
         \includegraphics[width=0.116\textwidth]{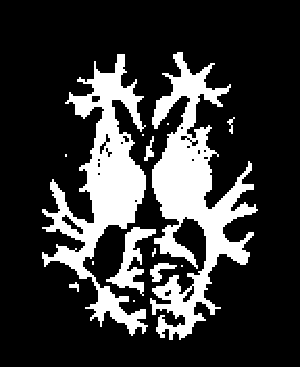}&
         \includegraphics[width=0.116\textwidth]{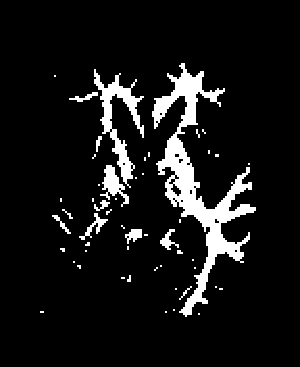}&
         \includegraphics[width=0.116\textwidth]{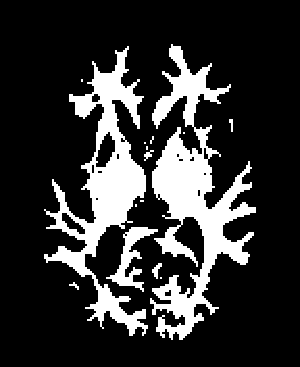}\\
         
         Domain 4&
         \includegraphics[width=0.116\textwidth]{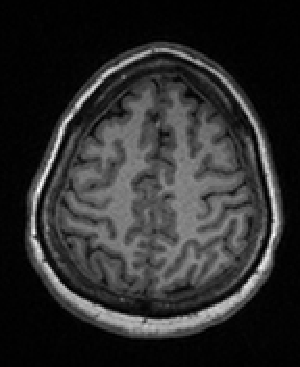}&
         \includegraphics[width=0.116\textwidth]{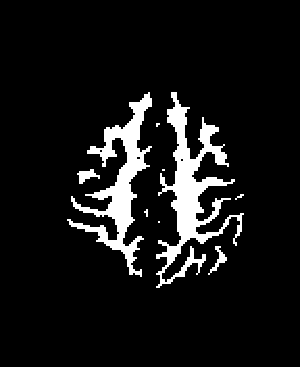}&
         \includegraphics[width=0.116\textwidth]{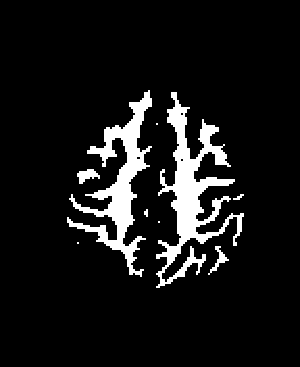}&
         \includegraphics[width=0.116\textwidth]{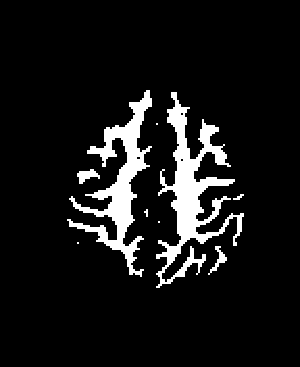}&
         \includegraphics[width=0.116\textwidth]{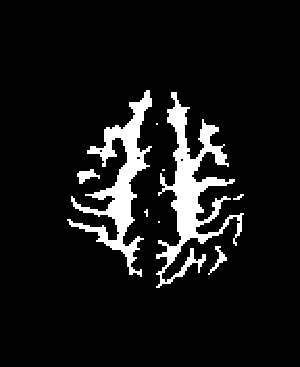}&
         \includegraphics[width=0.116\textwidth]{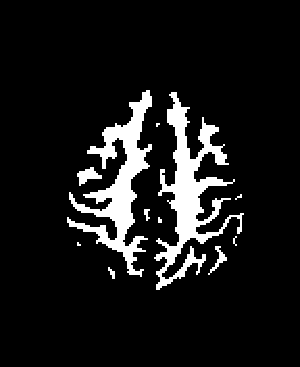}&
         \includegraphics[width=0.116\textwidth]{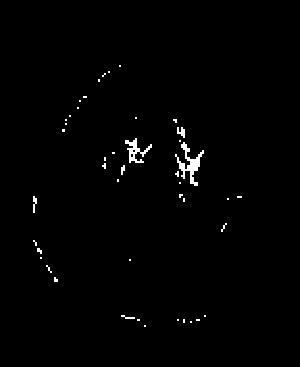}&
         \includegraphics[width=0.116\textwidth]{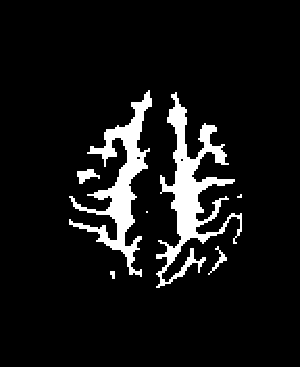}\\

     \end{tabular}
     }
      \caption{The results of the white matter segmentation network are presented, with each column representing the respective harmonization method applied to the input image.}
   \label{fig:downstream}
 \end{figure*}

\begin{table}[]
\setlength{\tabcolsep}{3pt}
\centering
{\footnotesize
   \begin{tabular}{ccccccccc}
      
\hline
    & IoU ($\uparrow$) & Domain 1 & Domain 2 & Domain 3 & Domain 4   \\
    \hline \hline
    &Source        & 0.912          & 0.845          & 0.947          & 0.938   \\
    &BlindHarmony (Ours)  & \textbf{0.922} & \textbf{0.878} & \textbf{0.947} & \textbf{0.938} \\
    &HM            & 0.863          & 0.854          & 0.911          & 0.894    \\
    &SSIMH\cite{guan2022fast}         & 0.777          & 0.752          & 0.862          & 0.860   \\
	&CycleGAN$_{each}$\cite{zhu2017unpaired}      & 0.306          & 0.408          & 0.394          & 0.299     \\
	&U-net$_{each}$\cite{ronneberger2015u}         & 0.785          & 0.797          & 0.829          & 0.870   \\
	\hline
    \end{tabular}}
   \caption{The IOU values of the results of the segmentation network and label mask are reported. The case using BlindHarmony illustrated the best results when compared to other harmonization methods.}
   \label{table:downstream}
\end{table}

To further assess the impact of harmonization, we conducted an evaluation of the downstream task of white matter segmentation. For this task, a white matter segmentation network was trained to generate masks of white matter from given the T$_1$-weighted images. Notably, the network was solely trained on the target domain dataset, allowing us to measure harmonization performance through segmentation results with harmonized images as inputs. The white matter labels were generated using FSL FAST \cite{zhang2001segmentation}, and we adopted U-Net as the neural network architecture. Figure \ref{fig:downstream} shows the white matter segmentation results for each harmonization method. Remarkably, the outcomes demonstrate that BlindHarmony enhances segmentation performance, thus successfully harmonizing source domain images to the target domain more effectively than the other methods. For quantitative analysis, we computed the intersection over union (IoU) values between the predicted white matter masks and the label masks (Table \ref{table:downstream}). The IoU values were higher for BlindHarmony, further confirming its superior harmonization performance. 

It is important to note that the dataset size used for U-net and CycleGAN training was smaller than that used for the flow model training due to the requirement of multiple domains for training. In addition, the U-net dataset size was even smaller due to the requirement of paired data. (\eg, Flow model training: 75,240 slices, Domain 1 CycleGAN training: 75,240 slices from the target domain and 17,400 slices from the source domain, Domain 1 U-net training: 5,100 slices; See Supplementary materials) Moreover, the construction of paired datasets required the registration of the source domain image to the target domain image using the FSL FLIRT function. Despite these efforts, there may still exist misregistration in the source domain-target domain pairs, which can negatively affect the training procedure and result in slightly blurred images produced by U-Net. These may be the reason for the inferior performance of CycleGAN and U-Net compared to their application for simulated source domains.

\section{Discussion}
In this work, we proposed BlindHarmony, a blind harmonization framework for harmonizing MR images from the source domain to the target domain. This framework does not require source domain data during training and can be applied to unseen source domain images. The flow-based prior distribution network is trained, and the harmonized images are optimized using a distance between the source domain image and the sampled image while using regularization based on the magnitude of the latent vector.

The fact that BlindHarmony does not require source domain data during training is beneficial when applying the neural network to an unknown dataset or a dataset that does not have sufficient data for training. For example, a deep learning-based API provider that utilizes a network trained on a certain dataset may not know the source domain information. In this scenario, our framework can be used as an excellent initial approach to harmonize data until sufficient data is collected for other methods.

In the real-world dataset evaluation involving the downstream task of white matter segmentation, BlindHarmony demonstrated superior performance compared to CycleGAN and U-Net, which were explicitly trained on specific source domains. While U-Net is good at image translation, it may miss fine structures. As for CycleGAN, it was originally designed to transfer style from source to target domains. In this harmonization case, the fine structure in MR images can also be considered as a style from the perspective of CycleGAN, leading to structural distortion in CycleGAN harmonization. Unlike image-to-image translation neural networks, BlindHarmony can enforce the structural information of the source domain images while harmonizing the images. This unique feature prevents the introduction of structural distortions and ensures more accurate and reliable harmonization results.

\subsection{Limitations}
BlindHarmony incorporates iterative optimization in both the image and latent vector domains. In order to reduce the computational burden of calculating the gradient of the network parameters, we have ignored the determinant term in Equation \ref{eq:harmodel2}. Although this may compromise mathematical rigor, we believe that this simplification makes calculations easier and reduces processing time, making BlindHarmony more advantageous for practical use.

\begin{table}[]
\centering
      {\footnotesize
   \begin{tabular}{ccc}
      \hline
       Gamma 1.5                                &PSNR($\uparrow$) \\
                                       \hline \hline
      Source                           & 22.6          \\
      BlindHarmony ($\beta_1 = 1000$) & 25.7           \\
      BlindHarmony ($\beta_1 = 500$)  &  \textbf{26.8} \\
      HM                               & 26.5          \\
      SSIMH                             & 26.3         \\ \hline
      \end{tabular}}
      \caption{In the case of a simulated source domain with a Gamma transformation of 1.5, BlindHarmony exhibited inferior results when using predefined hyperparameters. However, fine-tuning the hyperparameters led to improvements in the harmonization outcomes.}
   \label{table:fail}
\end{table}

Additionally, it should be noted that BlindHarmony may not be applicable in every domain. If the distance between two images defined in Equation \ref{eq:dis} cannot capture the relationship between images of different domains (\eg, multi-contrast: T$_1$-weighted and T$_2$-weighted images), the optimization may fail and lead to poor results. For example, when applying an extreme contrast variation case such as simulated source domain data with Gamma transformation with a power of 1.5, BlindHarmony showed inferior results to conventional methods. (Table. \ref{table:fail}) Therefore, it is important to carefully consider the suitability of BlindHarmony for different applications. 

However, fine-tuning hyperparameters for each source domain may give improved results. As shown in Table \ref{table:fail}, the PSNR value increased when we changed the hyperparameter $\beta_1$ from 1000 to 500. Tuning these hyperparameters for each source domain may give a successful application of our approach in various scenarios, providing a highly adaptable and versatile solution. Future work may include optimizing these parameters using a transfer learning approach, as demonstrated in \cite{wei2022deep}.

\section{Conclusion}
In this study, we propose BlindHarmony, a flow-bassed blind harmonization method for MR images. Unlike other existing harmonization methods, our network is trained exclusively on the target domain dataset and can be applied to previously unseen domain images. The flow model is trained only on the target domain data, and the harmonized image is optimized to have a correlation with the source domain image while maintaining a high probability of the flow model. Both simulated and real-world datasets showed that our method achieves acceptable results. Our study demonstrates the feasibility of blind harmonization, providing an advantage in scenarios where access to source domain data is limited or unavailable.\\

\noindent \textbf{Acknowledgements} This work was supported by the funding agencies of the Republic of Korea (NRF- 2022R1A4A1030579, NRF-2021R1A2B5B03002783, 21NPSS-C163415-01, and IITP-2023-RS-2023-00256081), Radisen Co. Ltd. and INMC and IOER of Seoul National University. 

{\small
\bibliographystyle{ieee_fullname}
\bibliography{egbib}
}

\end{document}


\title{Supplementary materials for
   
BlindHarmony: ``Blind'' Harmonization for MR Images via Flow model}

\maketitle
\ificcvfinal\thispagestyle{empty}\fi

\section{Dataset description}
The BlindHarmony framework proposed in this work was trained and evaluated using the OASIS3 \cite{lamontagne2019oasis}, employing a target domain consisting of images acquired with a Siemens TIM Trio 3T MR scanner. Meanwhile, as a source domain, images obtained from four other scanners were utilized, including the Siemens Magnetom Vida 3T MR scanner (Domain 1), the Siemens Vision 1.5T scanner (Domain 2), the Siemens BioGraph mMR PET-MR 3T scanner (Domain 3), and the Siemens TIM Trio 3T MR scanner (Domain 4). Domain 4 scanner shared the same scanner version with the target domain scanner. All images were resampled to a uniform resolution of 1.2$\times$1.2$\times$1.2 mm and normalized using min-max normalization at the slice level. Table \ref{table:dataset} provides detailed information regarding the acquisition parameters.

\begin{table*}[]
   \begin{center}
   \begin{tabular}{cccccc}
   \hline
              & Target domain & Domain 1 & Domain 2 & Domain 3 & Domain 4 \\
              \hline\hline
   Manufacturer                & Siemens  & Siemens & Siemens & Siemens & Siemens \\
   scanner version             & TIM Trio & Magnetom Vida & Vision & BioGraph mMR & TIM Trio \\
   Magnetic field strength (T)   & 3 &3 &1.5 & 3 & 3\\
   Matrix size                & 176$\times$256$\times$256 &176$\times$240$\times$256 & 128$\times$256$\times$256 & 176$\times$240$\times$256 & 176$\times$256$\times$256\\
   resolution (mm)                  & 1$\times$1$\times$1 &1.2$\times$1.05$\times$1.05 & 1.25$\times$1$\times$1 & 1.2$\times$1.05$\times$1.05 & 1$\times$1$\times$1\\
   TR/TI (s)     &2.4/1 & 2.3/unknown & 9.7/unknown & 2.3/0.9 & 2.4/1 \\
   TE (ms)      &3.2 & 3.0 & 4 & 3.0 & 3.2 \\
   Flip angle ($^{\circ}$)     &8 & 9 & 10 & 9 & 8 \\
   Total number of sessions     & 1568 & 378 & 620 & 879 & 625 \\
    \hline
   \end{tabular}
   
\end{center}
   \caption{Scan parameters of domains in OASIS3 dataset is illustrated. }
   \label{table:dataset}
\end{table*}

\section{Training details}
\subsection{Flow model}
\begin{figure}
   \begin{center}
      \includegraphics[width= 0.5\textwidth]{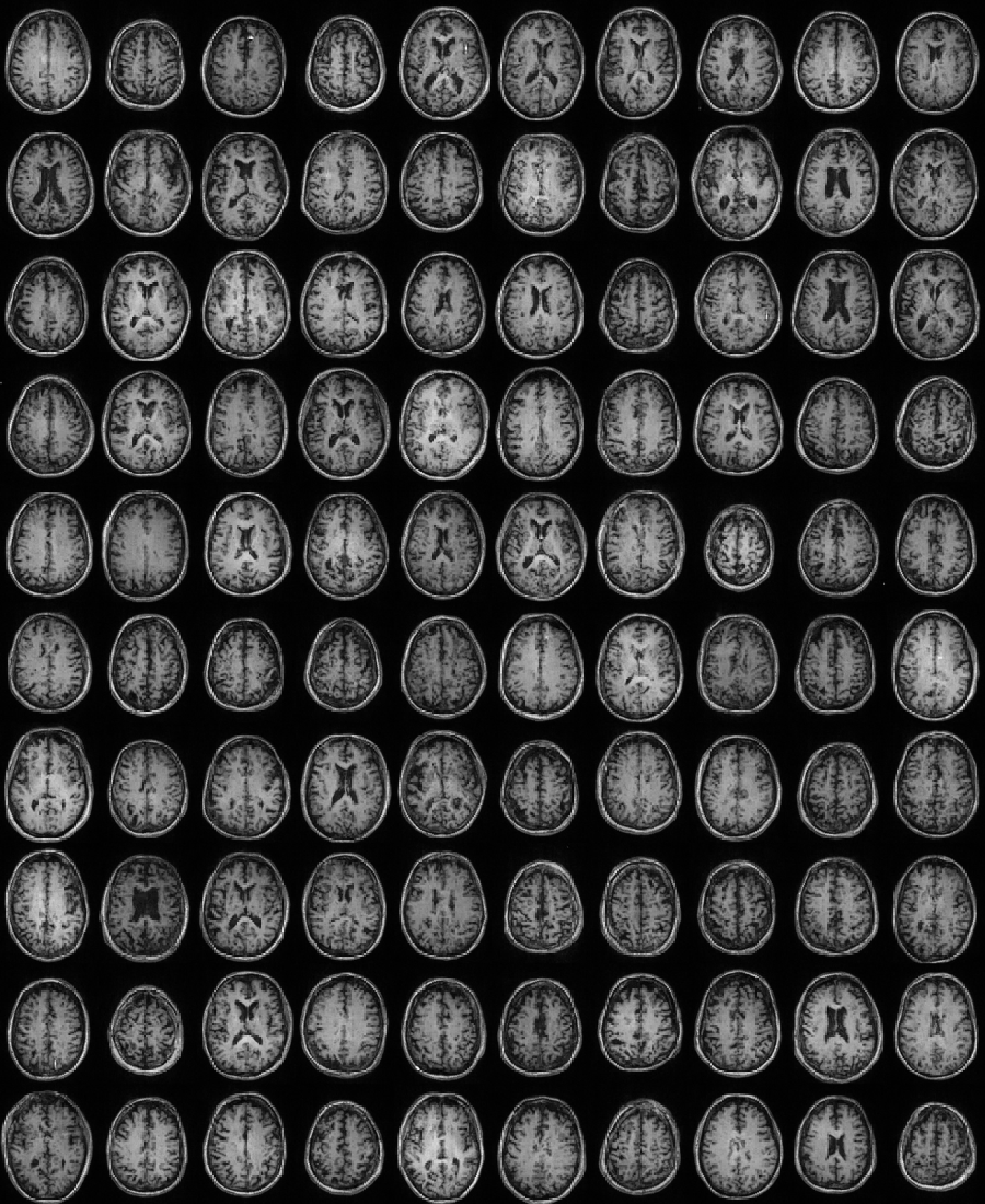}
   \end{center}
      \caption{Example images of the trained flow model.}
   \label{fig:ex}
   \end{figure}

For our experiments, we employed the Neural Spline Flow (NSF) architecture \cite{durkan2019neural} with rational quadratic (RQ) spline coupling layers. The majority of the hyperparameters for the network were set to the same values used in the original NSF paper for experiments on the ImageNet dataset. Specifically, the tail bound B and the number of bins K for the RQ spline coupling layers were set to B = 3 and K = 8, respectively. A multiscale architecture similar to that of Glow was utilized, with each layer of the network consisting of 7 transformation steps, including an actnorm layer, an invertible 1 $\times$ 1 convolution, an RQ spline coupling transform, and another 1 $\times$ 1 convolution. The network itself comprises 4 layers, which results in a total of 28 coupling transformation steps. Additionally, 3 residual blocks and batch normalization layers are included in the subnetworks parameterizing the RQ splines. An Adam optimizer with an initial learning rate of 0.0005 and cosine annealing of the learning rate was used to iteratively optimize the parameters up to 20K steps. Two separate models were trained, one for simulated dataset validation and one for real-world dataset validation. The same hyperparameters were assigned to both models except for the dataset composition, with the simulated evaluation dataset consisting of 76775/775/800 slices for training/validation/testing, and the real-world evaluation dataset consisting of 75240/760/1000 slices for training/validation/testing. An example of a sampled image of the flow model is illustrated in Figure \ref{fig:ex}.

\begin{table*}[]
   \begin{center}
      
   \begin{tabular}{cccccc}
   \hline
               & Domain 1 & Domain 2 & Domain 3 & Domain 4 \\
               \hline \hline
   U-net (slices)      & 5100/1000 & 9750/600 & 25150/1000 & 19350/1000 \\
   CylceGAN (slices)   & 17400/1000 & 29000/600 & 42100/1000 & 29050/1000 \\
   \hline
   \end{tabular}
   
\end{center}
   \caption{The number of training/test datasets for each domain. }
   \label{table:training}
\end{table*}

\subsection{U-Net}
The U-Net \cite{ronneberger2015u} architecture includes 4 Down blocks and 4 Up blocks, each block consisting of two sequences of convolution layer, batch normalization, and ReLU activation. The Down blocks utilize maxpooling, while the Up blocks use bilinear upsampling. The U-Net also utilizes skip connections. The training was done with 120 epochs using L1 loss and the Adam optimizer with a learning rate of 0.001. For the simulated dataset evaluation, the same dataset composition used in training the flow model was utilized, with a composition of train/val/test = 76775/775/800 slices. For the real-world dataset application, the validation step was dropped to increase the number of training data. The number of datasets varies by domain, as described in Table \ref{table:training}.

\subsection{CycleGAN}
The generator in CycleGAN \cite{zhu2017unpaired} consists of 2 convolution layers with instance normalization and ReLU activation, 9 residual blocks, and 3 convolution layers with instance normalization and ReLU activation. Each residual block includes a residual connection of 2 convolution layers with instance normalization and ReLU activation. The discriminator consists of 5 convolutional layers, 4 leaky ReLU, and 4 instance normalization. The CycleGAN uses identity loss, cycle loss, and adversarial loss, and training was done with 40 epochs using the Adam optimizer with a learning rate of 0.0002. For the simulated dataset evaluation, the same dataset composition used in training the flow model was utilized, with a composition of train/test = 76775/775/800 slices. For the real-world dataset application, the validation step was dropped as in the U-Net cases, and the number of target domain images was train/test = 75240/1000 slices, while the number of source domain images varied by domain, as described in Table \ref{table:training}.

\section{Ablation study}
We compared the PSNR and SSIM values of a simulated dataset to a baseline by varying one of the hyperparameters $\alpha$, $\beta_1$, and $\beta_2$.

\begin{table}[]
   \begin{center}
      
   \begin{tabular}{ccccc}
   \hline
               PSNR & Exp & Log & Gamma 0.7 \\
               \hline \hline
   Baseline      & 29.6 & 28.8 & 27.4 \\
   $\alpha=0$  & 29.5 & 28.5 & 27.2  \\
   $\beta_1=0$  & 18.5 & 18.5 & 18.6 \\
   $\beta_2=0$  & 29.6 & 28.8 & 27.4 \\
   \hline
   \\

   \hline
   SSIM & Exp & Log & Gamma 0.7 \\
   \hline\hline
   Baseline      & 0.985 & 0.978 & 0.969 \\
   $\alpha=0$  & 0.984 & 0.977 & 0.968  \\
   $\beta_1=0$  & 0.693 & 0.695 & 0.696 \\
   $\beta_2=0$  & 0.985 & 0.978 & 0.969 \\
   \hline
   \end{tabular}
   
\end{center}
   \caption{The ablation study. }
   \label{table:abl}
\end{table}

\section{Illustration for Eq. 1 and Eq. 2}
\begin{figure}
   \includegraphics[width=\linewidth]{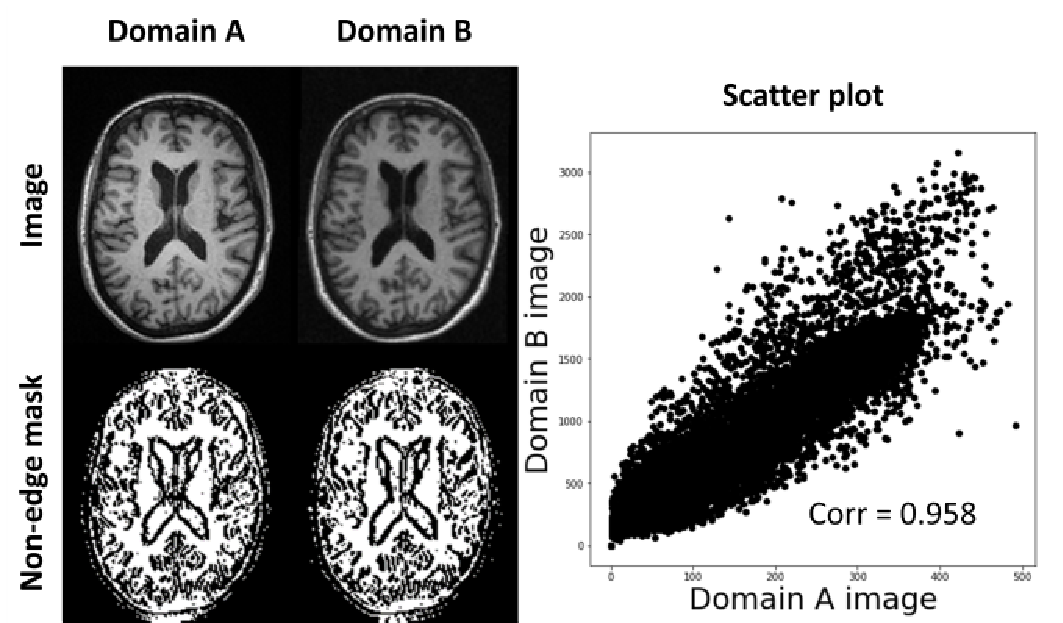}
      \caption{Visual illustration for Eq. 1 and Eq. 2.}
\label{fig:distance}
\end{figure}
When harmonizing two T$_1$ weighted images from different scanners, the images are primarily created by the properties of the brain (\eg, T1, proton density) while scanner-specific differences are mostly spatially slow-varying patterns (\eg, center-brightening from B1 inhomogeneity). Consequently, a high spatial correlation (Eq. 1) exists between the images, and their edges are likely to coincide (Eq. 2). These can be observed in Fig. \ref{fig:distance}. The non-edge masks of the images from the two different domains exhibit a high coincidence, and the scatter plot demonstrates a strong correlation between the two images. If these conditions are not met, it may create a failure case (e.g., harmonizing T$_1$ and T$_2$ weighted images).

\section{Visual examples}
\begin{figure*}
\setlength{\tabcolsep}{0pt}
\centering
   {\scriptsize
\begin{tabular}{C{1.2cm}C{1.8cm}C{1.8cm}C{1.8cm}C{0.1cm}C{1.8cm}C{1.8cm}C{1.8cm}C{0.1cm}C{1.8cm}C{1.8cm}C{1.8cm}}
&Source & Target & BlindHarmony &  &Source & Target & BlindHarmony& &Source & Target & BlindHarmony\\
         Exp&
         \includegraphics[width=0.1\textwidth]{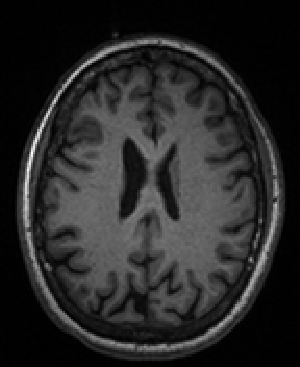}&
         \includegraphics[width=0.1\textwidth]{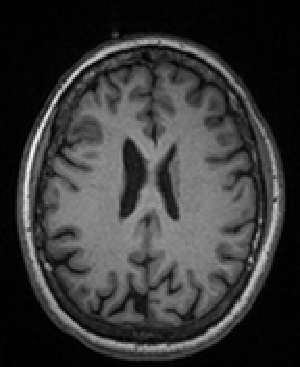}&
         \includegraphics[width=0.1\textwidth]{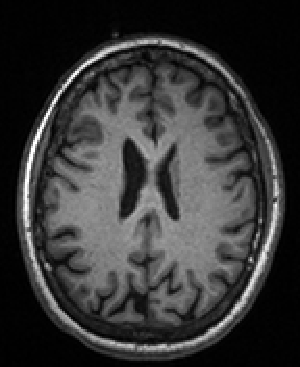}&
         &
         \includegraphics[width=0.1\textwidth]{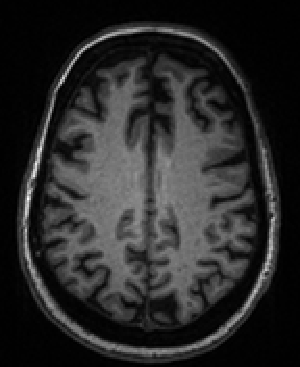}&
         \includegraphics[width=0.1\textwidth]{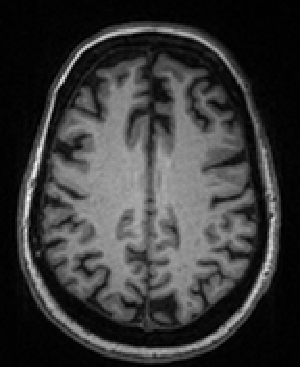}&
         \includegraphics[width=0.1\textwidth]{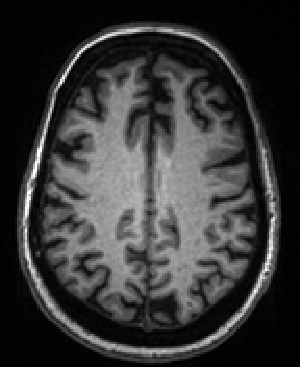}&
         &
         \includegraphics[width=0.1\textwidth]{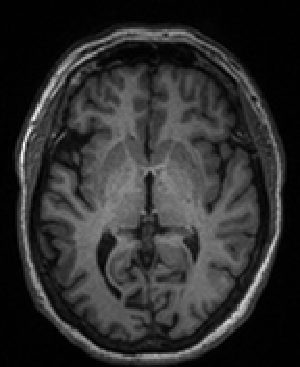}&
         \includegraphics[width=0.1\textwidth]{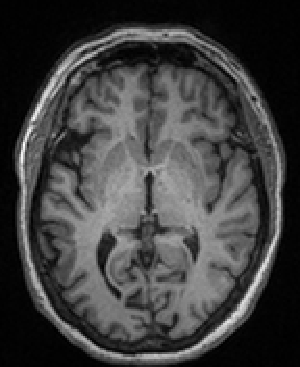}&
         \includegraphics[width=0.1\textwidth]{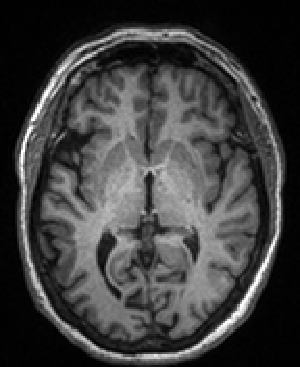}\\
         \\
         Log&
         \includegraphics[width=0.1\textwidth]{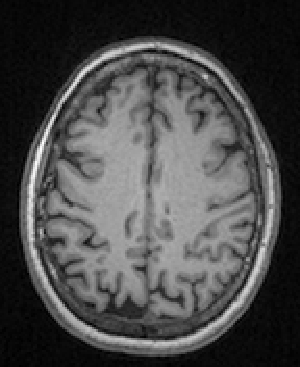}&
         \includegraphics[width=0.1\textwidth]{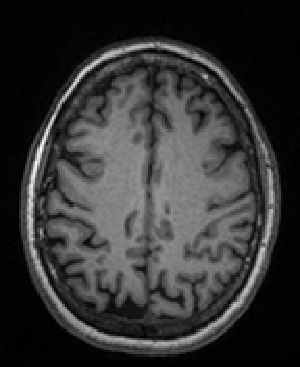}&
         \includegraphics[width=0.1\textwidth]{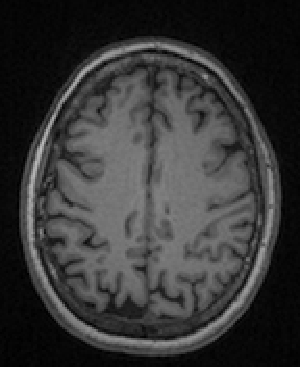}&
         &
         \includegraphics[width=0.1\textwidth]{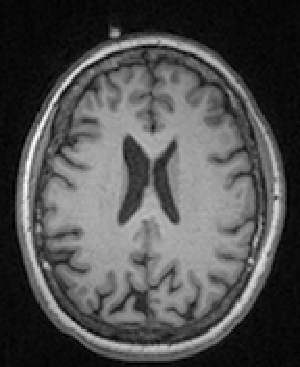}&
         \includegraphics[width=0.1\textwidth]{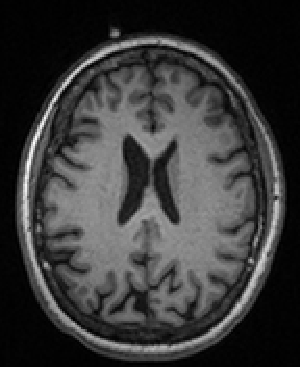}&
         \includegraphics[width=0.1\textwidth]{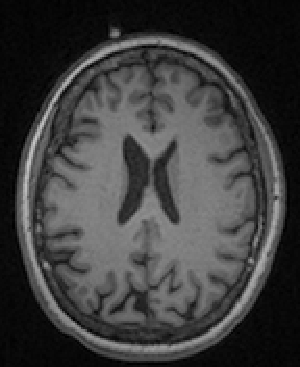}&
         &
         \includegraphics[width=0.1\textwidth]{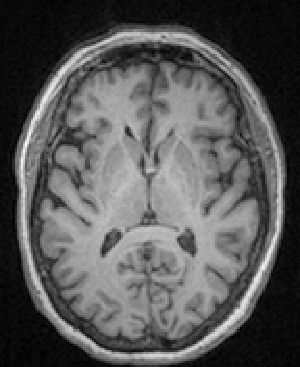}&
         \includegraphics[width=0.1\textwidth]{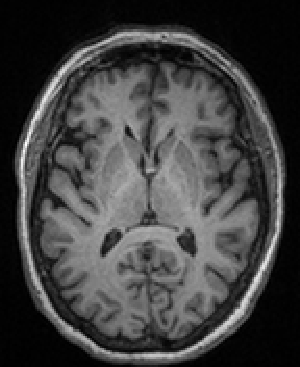}&
         \includegraphics[width=0.1\textwidth]{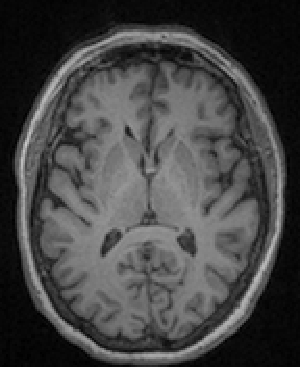}\\
         \\
         Gamma 0.7&
         \includegraphics[width=0.1\textwidth]{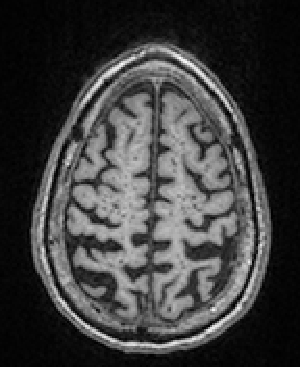}&
         \includegraphics[width=0.1\textwidth]{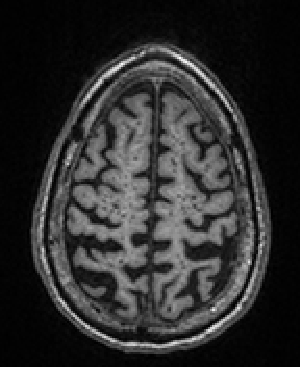}&
         \includegraphics[width=0.1\textwidth]{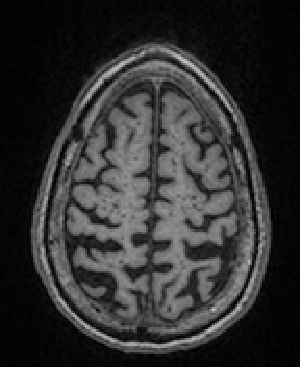}&
         &
         \includegraphics[width=0.1\textwidth]{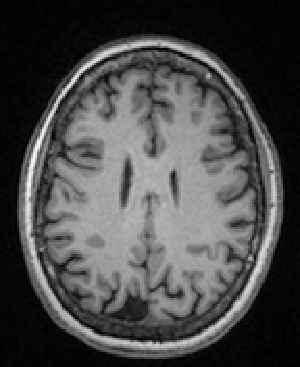}&
         \includegraphics[width=0.1\textwidth]{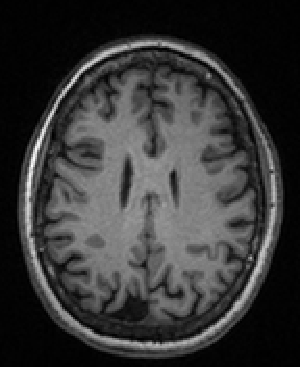}&
         \includegraphics[width=0.1\textwidth]{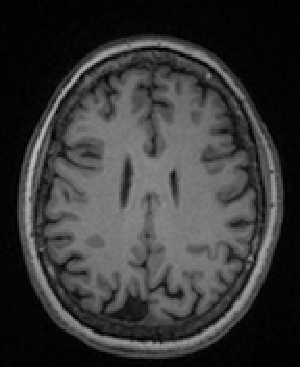}&
         &
         \includegraphics[width=0.1\textwidth]{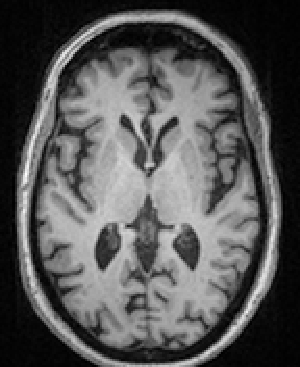}&
         \includegraphics[width=0.1\textwidth]{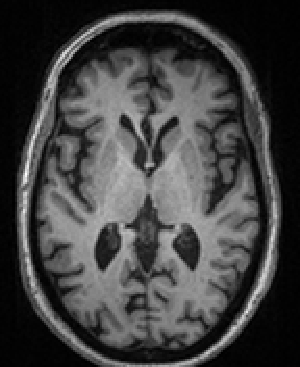}&
         \includegraphics[width=0.1\textwidth]{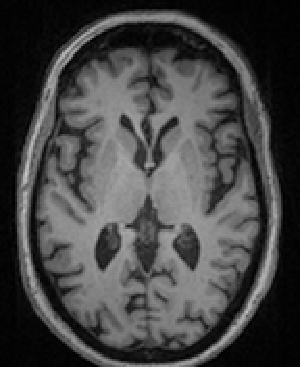}\\
     \end{tabular}
     }
     \caption{Example images of BlindHarmony application of simulated source domain images.}
   \label{fig:retro}
 \end{figure*}

\begin{figure*}
\setlength{\tabcolsep}{0pt}
\centering
   {\scriptsize
\begin{tabular}{C{1.2cm}C{1.8cm}C{1.8cm}C{1.8cm}C{0.1cm}C{1.8cm}C{1.8cm}C{1.8cm}C{0.1cm}C{1.8cm}C{1.8cm}C{1.8cm}}
&Source & Target & BlindHarmony &  &Source & Target & BlindHarmony& &Source & Target & BlindHarmony\\

        Domain 1&
         \includegraphics[width=0.1\textwidth]{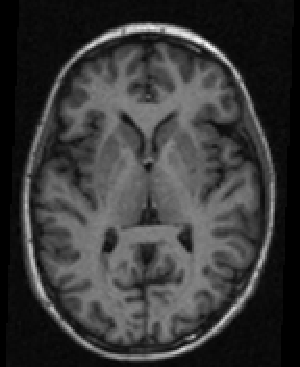}&
         \includegraphics[width=0.1\textwidth]{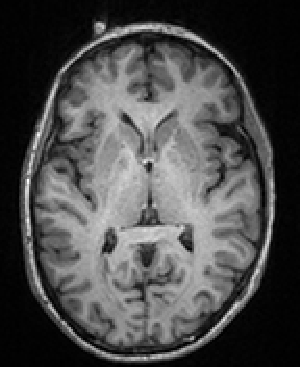}&
         \includegraphics[width=0.1\textwidth]{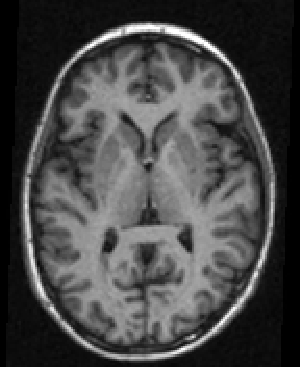}&
         &
         \includegraphics[width=0.1\textwidth]{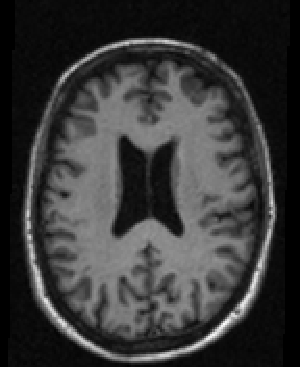}&
         \includegraphics[width=0.1\textwidth]{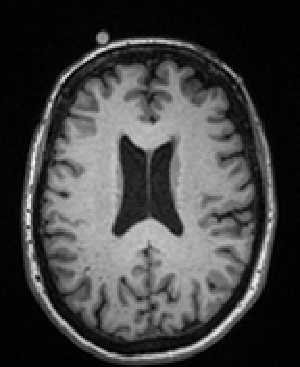}&
         \includegraphics[width=0.1\textwidth]{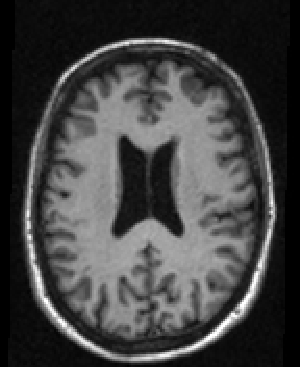}&
         &
         \includegraphics[width=0.1\textwidth]{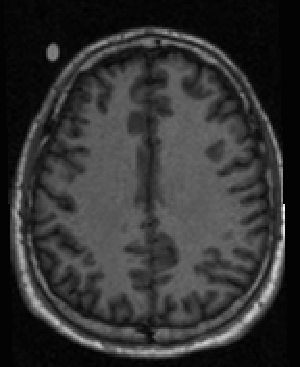}&
         \includegraphics[width=0.1\textwidth]{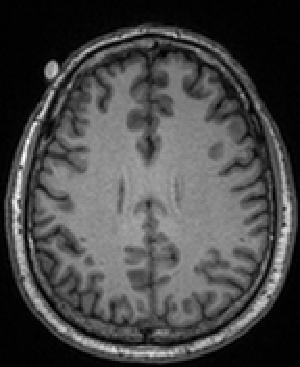}&
         \includegraphics[width=0.1\textwidth]{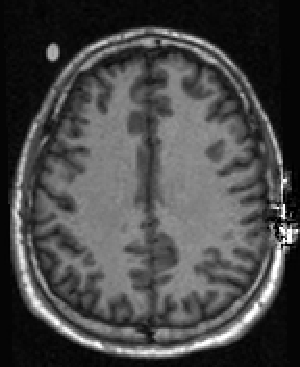}\\
         \\
         Domain 2&
         \includegraphics[width=0.1\textwidth]{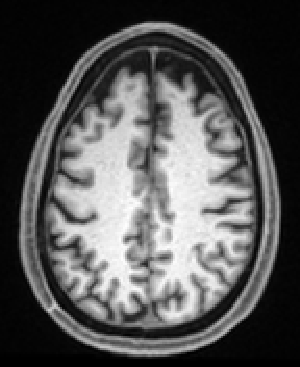}&
         \includegraphics[width=0.1\textwidth]{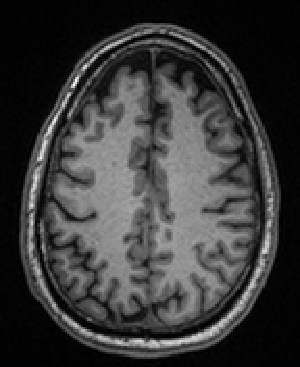}&
         \includegraphics[width=0.1\textwidth]{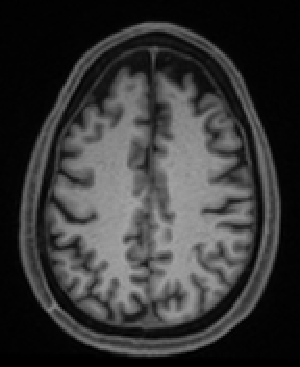}&
         &
         \includegraphics[width=0.1\textwidth]{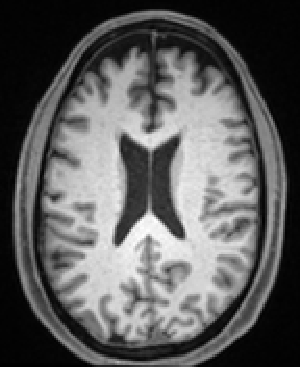}&
         \includegraphics[width=0.1\textwidth]{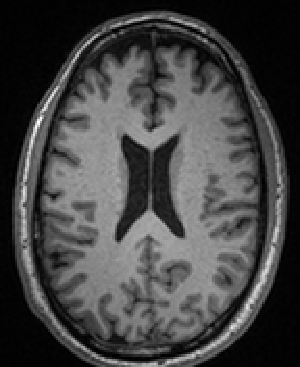}&
         \includegraphics[width=0.1\textwidth]{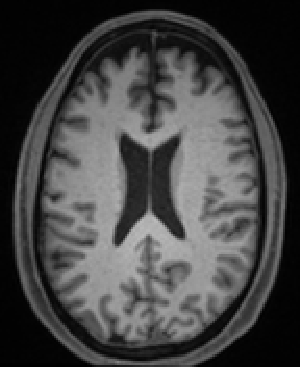}&
         &
         \includegraphics[width=0.1\textwidth]{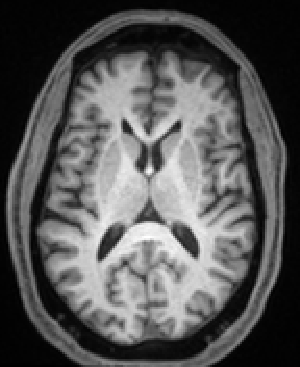}&
         \includegraphics[width=0.1\textwidth]{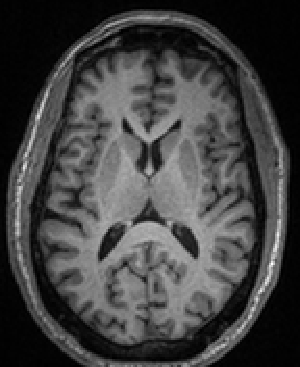}&
         \includegraphics[width=0.1\textwidth]{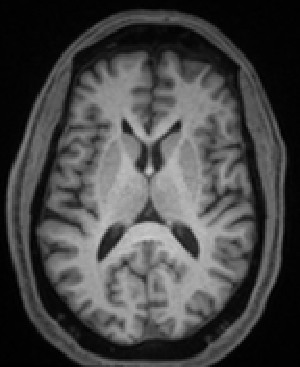}\\
         \\
         Domain 3&
         \includegraphics[width=0.1\textwidth]{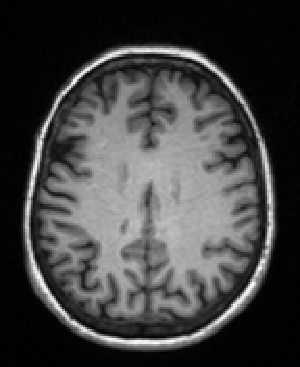}&
         \includegraphics[width=0.1\textwidth]{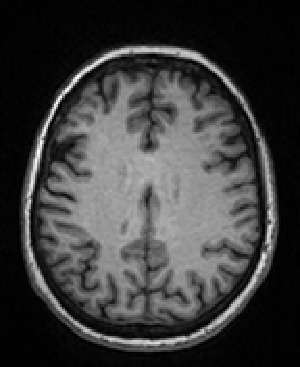}&
         \includegraphics[width=0.1\textwidth]{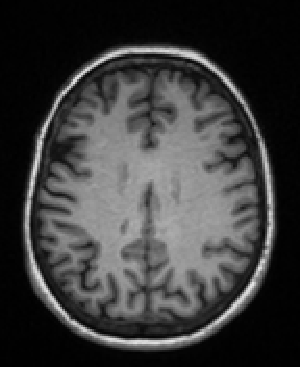}&
         &
         \includegraphics[width=0.1\textwidth]{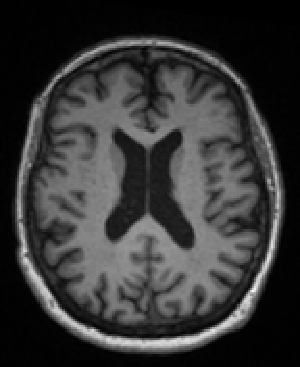}&
         \includegraphics[width=0.1\textwidth]{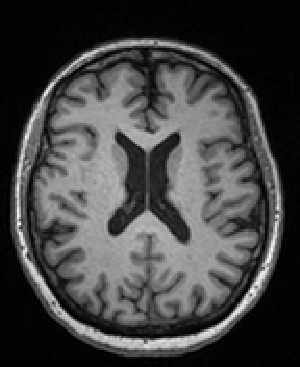}&
         \includegraphics[width=0.1\textwidth]{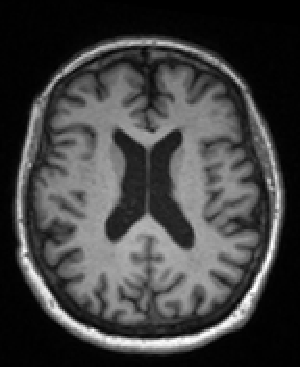}&
         &
         \includegraphics[width=0.1\textwidth]{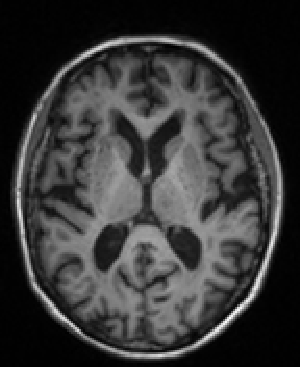}&
         \includegraphics[width=0.1\textwidth]{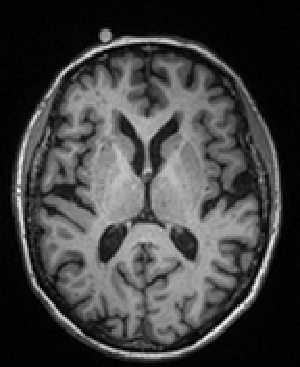}&
         \includegraphics[width=0.1\textwidth]{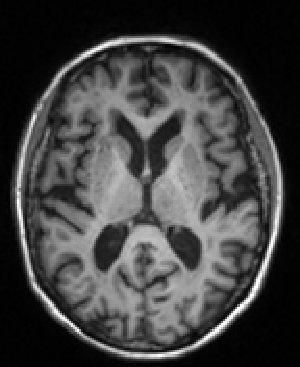}\\
      \\
         Domain 4&
         \includegraphics[width=0.1\textwidth]{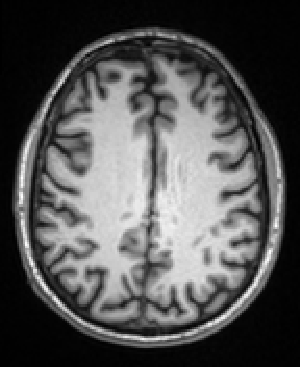}&
         \includegraphics[width=0.1\textwidth]{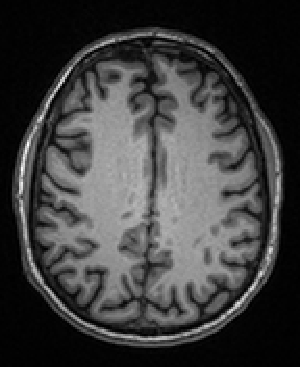}&
         \includegraphics[width=0.1\textwidth]{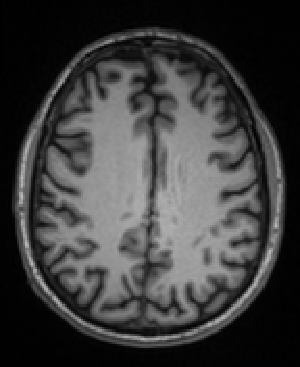}&
         &
         \includegraphics[width=0.1\textwidth]{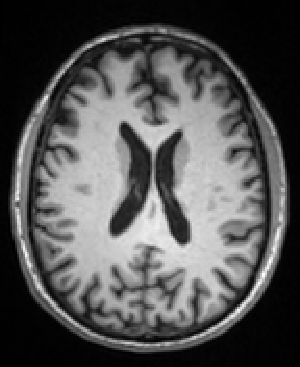}&
         \includegraphics[width=0.1\textwidth]{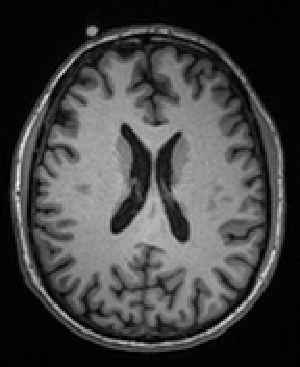}&
         \includegraphics[width=0.1\textwidth]{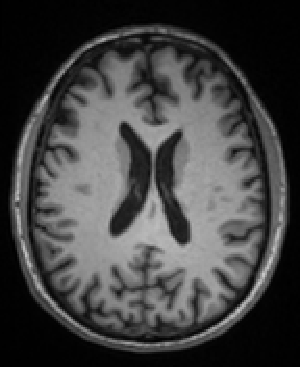}&
         &
         \includegraphics[width=0.1\textwidth]{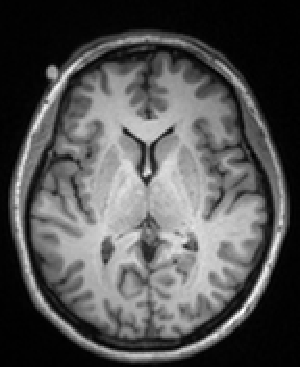}&
         \includegraphics[width=0.1\textwidth]{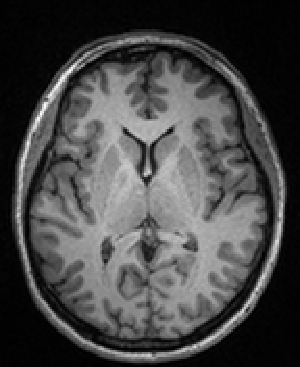}&
         \includegraphics[width=0.1\textwidth]{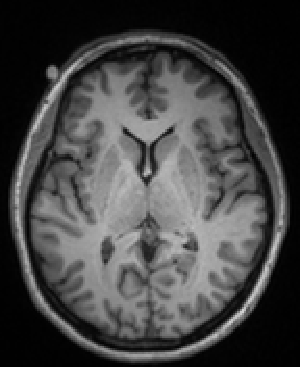}\\
     \end{tabular}
     }
     \caption{Example images of BlindHarmony application of real-world data images.}
   \label{fig:pro}
 \end{figure*}

\begin{figure*}
\setlength{\tabcolsep}{0pt}
\centering
   {\scriptsize
\begin{tabular}{C{1.2cm}C{2cm}C{2cm}C{2cm}C{2cm}C{0.1cm}C{2cm}C{2cm}C{2cm}C{2cm}}
&Source & Label & No harmonization & BlindHarmony&  &Source & Label & No harmonization & BlindHarmony\\

        Domain 1&
         \includegraphics[width=0.11\textwidth]{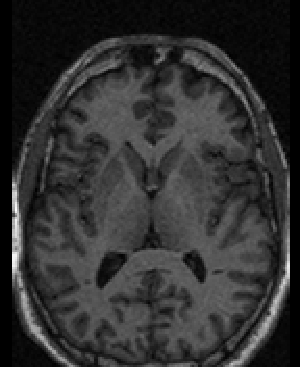}&
         \includegraphics[width=0.11\textwidth]{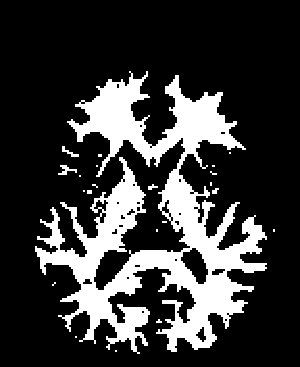}&
         \includegraphics[width=0.11\textwidth]{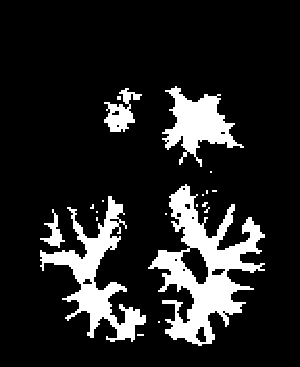}&
         \includegraphics[width=0.11\textwidth]{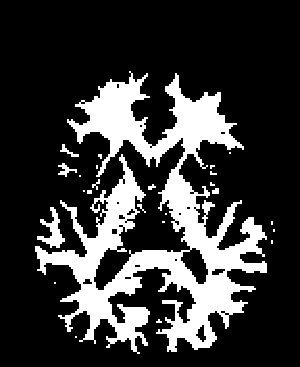}&
         &
         \includegraphics[width=0.11\textwidth]{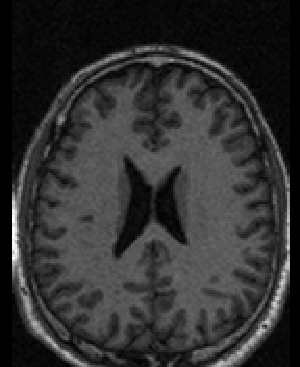}&
         \includegraphics[width=0.11\textwidth]{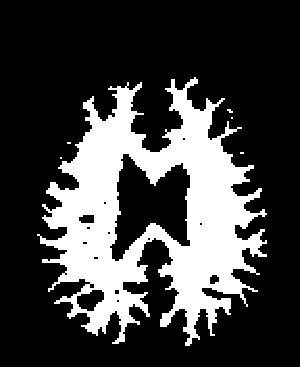}&
         \includegraphics[width=0.11\textwidth]{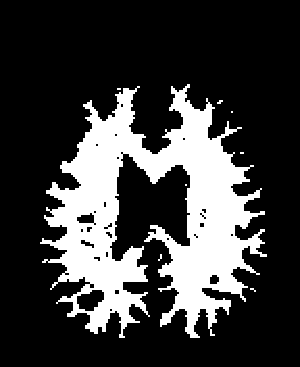}&
         \includegraphics[width=0.11\textwidth]{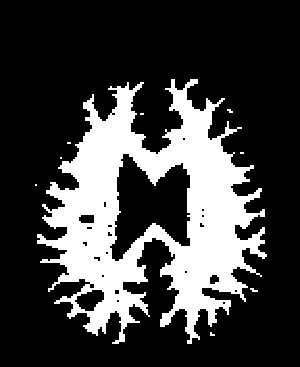}\\
         \\
         Domain 2&
         \includegraphics[width=0.11\textwidth]{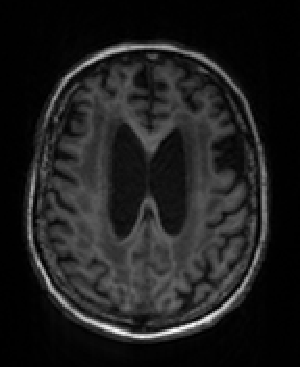}&
         \includegraphics[width=0.11\textwidth]{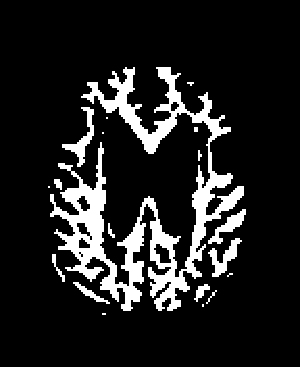}&
         \includegraphics[width=0.11\textwidth]{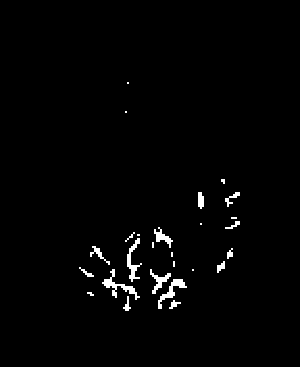}&
         \includegraphics[width=0.11\textwidth]{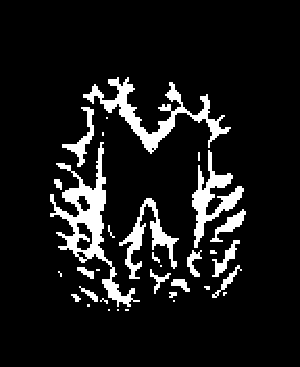}&
         &
         \includegraphics[width=0.11\textwidth]{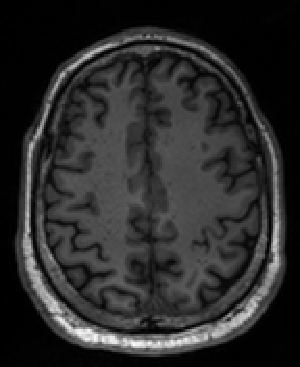}&
         \includegraphics[width=0.11\textwidth]{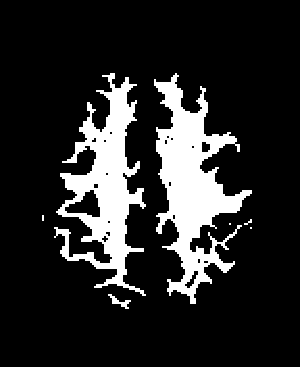}&
         \includegraphics[width=0.11\textwidth]{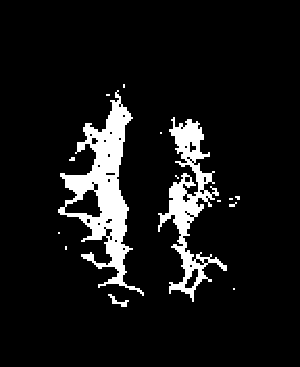}&
         \includegraphics[width=0.11\textwidth]{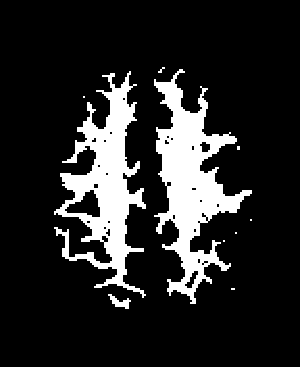}\\
         \\
         Domain 3&
         \includegraphics[width=0.11\textwidth]{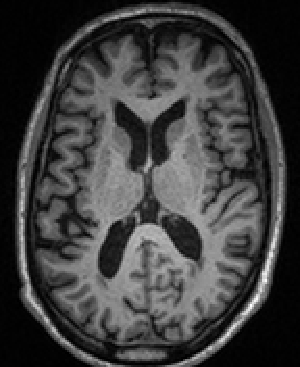}&
         \includegraphics[width=0.11\textwidth]{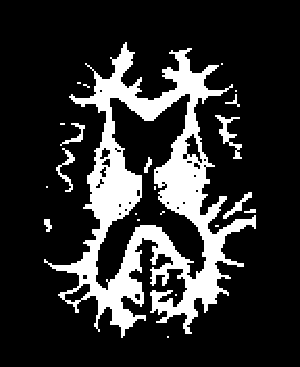}&
         \includegraphics[width=0.11\textwidth]{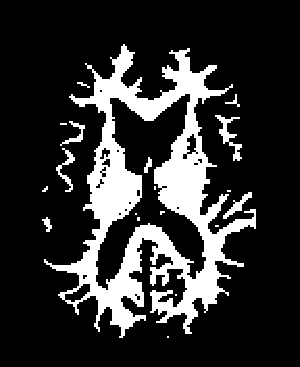}&
         \includegraphics[width=0.11\textwidth]{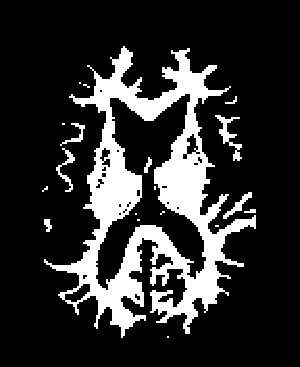}&
         &
         \includegraphics[width=0.11\textwidth]{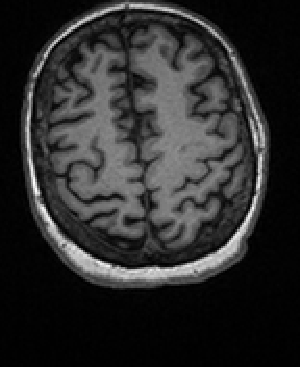}&
         \includegraphics[width=0.11\textwidth]{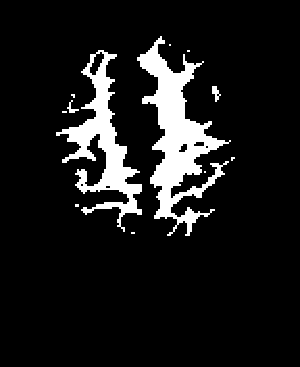}&
         \includegraphics[width=0.11\textwidth]{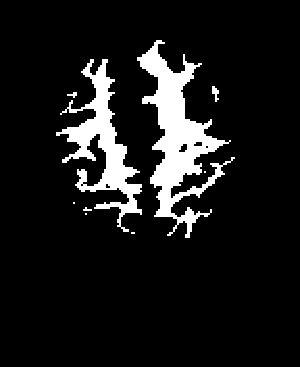}&
         \includegraphics[width=0.11\textwidth]{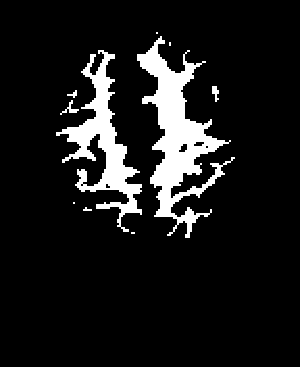}\\
      \\
         Domain 4&
         \includegraphics[width=0.11\textwidth]{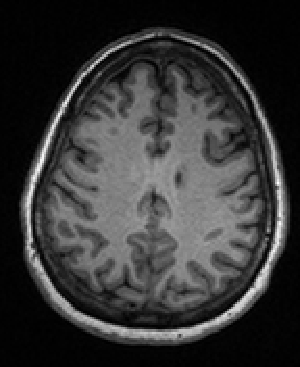}&
         \includegraphics[width=0.11\textwidth]{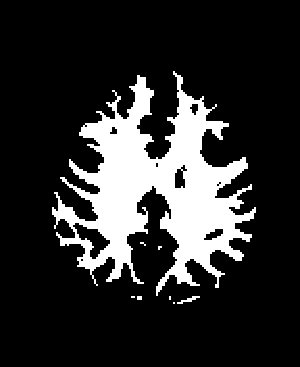}&
         \includegraphics[width=0.11\textwidth]{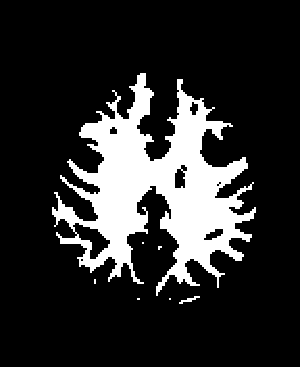}&
         \includegraphics[width=0.11\textwidth]{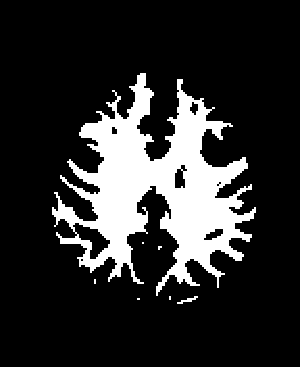}&
         &
         \includegraphics[width=0.11\textwidth]{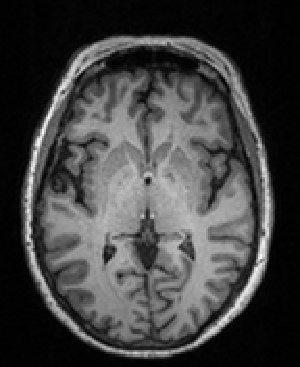}&
         \includegraphics[width=0.11\textwidth]{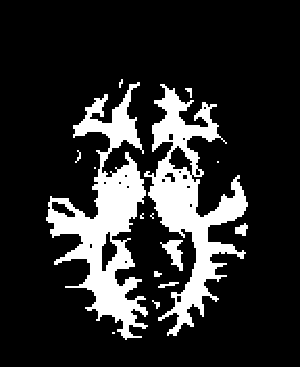}&
         \includegraphics[width=0.11\textwidth]{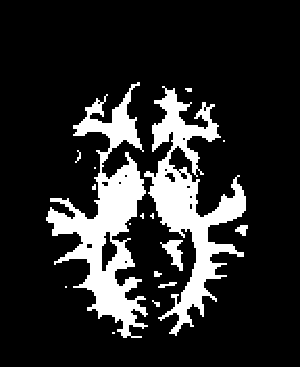}&
         \includegraphics[width=0.11\textwidth]{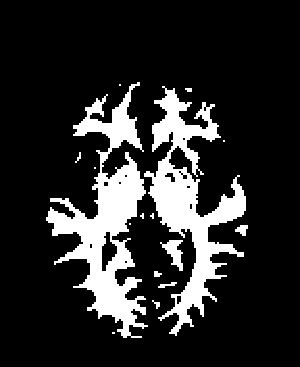}\\
     \end{tabular}
     }
     \caption{Example images of BlindHarmony application to the downstream task of white matter segmentation network.}
   \label{fig:task}
 \end{figure*}

Exemplary images of the simulated source domain dataset application, real-world data application, and the downstream task application are illustrated in Figures \ref{fig:retro}, \ref{fig:pro} and \ref{fig:task}

{\small
\bibliographystyle{ieee_fullname}
\bibliography{egbib}
}